\documentstyle[a4,12pt,axodraw]{article}

\def\pd{\partial}
\def\a{\alpha}
\def\b{\beta}
\def\dl{\delta}
\def\s{\sigma}

\def\eps{\epsilon}

\def\lam{\lambda}
\def\Lam{\Lambda}
\def\bg{{\bar g}}
\def\hg{{\hat g}}
\def\hh{{\hat h}}
\def\ta{{\tilde a}}
\def\tb{{\tilde b}}
\def\tc{{\tilde c}}
\def\tt{{\tilde t}}
\def\bnabla{{\bar \nabla}}
\def\hnabla{{\hat \nabla}}
\def\bR{{\bar R}}
\def\hR{{\hat R}}
\def\bBox{\stackrel{-}{\Box}}
\def\hBox{{\hat \Box}} 
\def\beps{{\bar \epsilon}}
\def\gm{\gamma}

\def\om{\omega}

\def\sq{\sqrt}
\def\e{\hbox{\large \it e}}
\def\half{\frac{1}{2}}
\def\fr{\frac}
\def\pp{\prime}
\def\arr{\rightarrow}
\def\lk{l \cdot k}
\def\bb{\begin{equation}}
\def\ee{\end{equation}}
\def\bba{\begin{eqnarray}}
\def\eea{\end{eqnarray}}

\begin{document}

\begin{titlepage}

\begin{tabbing}
   qqqqqqqqqqqqqqqqqqqqqqqqqqqqqqqqqqqqqqqqqqqqqq 
   \= qqqqqqqqqqqqq  \kill 
         \>  {\sc KEK-TH-656 }    \\
         \>       hep-th/9912098 \\
         \>  {\sc December, 1999} 
\end{tabbing}

\vspace{1.5cm}

\begin{center}
{\Large {\bf A Candidate for  
A Renormalizable and Diffeomorphism-Invariant  
4D Quantum Theory of Gravity}}
\end{center}

\vspace{1.5cm}

\centering{\sc Ken-ji Hamada\footnote{E-mail address : 
hamada@post.kek.jp} }

\vspace{1cm}

\begin{center}
{\it Institute of Particle and Nuclear Studies, \break 
High Energy Accelerator Research Organization (KEK),} \\ 
{\it Tsukuba, Ibaraki 305-0801, Japan}
\end{center} 

\begin{center}
$$
  \left(  \begin{array}{c} 
       \hbox{revised and extended version} \\ 
       \hbox{to appear in Prog. Theor. Phys.} 
          \end{array} 
  \right)
$$
\end{center}

\vspace{7mm}

\begin{abstract} 
We present evidence that there is a 4D model that satisfies the conditions 
of renormalizability and diffeomorphism invariance simultaneously at the 
2-loop level.     
The traceless mode is treated perturbatively, while the conformal mode 
can be managed exactly. 
The two conditions constrain the theory strongly and determine 
the measure of the gravitational field uniquely.  
Quantum corrections of the cosmological constant are computed 
in part to 3-loop diagrams. 
The method to remove the negative-metric states is also discussed from the 
viewpoint of diffeomorphism invariance in analogy to 
the $R_{\xi}$ gauge in spontaneously broken gauge theory.  
The model may be a candidate for a continuum version of 4D simplicial 
quantum geometry realized in recent numerical simulations.

%\noindent
%PACS: 04.60.-m 

%\noindent
%Keywords: 4Dquantum gravity; diffeomorphism invariance; renormalization; Wess-Zumino action.  
\end{abstract}
\end{titlepage}  

\section{Introduction}
\setcounter{equation}{0}
\noindent

 For a long time, many authors have been longing to construct 
a well-defined 4D quantum theory of gravity~\cite{d}--\cite{hs}. 
Renormalizability is a big problem that we must overcome in such 
a construction. For this reason, we must make a theory of this kind 
to be diffeomorphism invariant. 
Furthermore, there are physical problems, such as unitarity, 
and the correct manner to 
define the $S$-matrix in a fully fluctuating geometry. 
As a first step, it should be clarified whether there is a model that 
satisfies the two fundamental conditions of renormalizability 
and diffeomorphism invariance simultaneously. 
In this paper we give evidence that the model proposed in a previous 
paper~\cite{hs}  satisfies these two conditions. 
The problem of the $S$-matrix and  unitarity in our model is also 
discussed from the viewpoint of diffeomorphism invariance at the end of 
the paper. 

  Along another line, there have been developments in the dynamical 
triangulation approach~\cite{wei,bk,adj} to a 4-dimensional 
manifold~\cite{migdal,bbkptt}. 
Recent numerical simulations suggest that 4D simplicial quantum 
geometry with $U(1)$ gauge fields is likely to have a phase 
transition that is of higher order than first~\cite{bbkptt}.    
Naively, if there is a continuum limit of 4D simplicial quantum geometry, 
it should satisfy the conditions of renormalizability and 
diffeomorphism invariance. 
   
   Diffeomorphism invariant quantum theory is formally defined by 
the functional integration over the metric 
\bb
   Z= \int \frac{[g^{-1}dg]_g [df]_g}{\hbox{vol(diff.)}} 
          \exp [-I(f,g)] ~, 
              \label{zzz}
\ee
where $f$ is a matter field~\footnote{%%%%%%%%(footnote 2)%%%%%%
In this paper all second order fields, except the gravitational field,  
are referred to as ``matter fields".   
} %%%%%%%%%(end of footnote 2)%%%%%%
and $I$ is an invariant action.   
The measure of the metric is defined by the norm, 
\bb
      <d g, d g>_g = \int d^D x \sq{g} g^{\mu\nu}g^{\lam\s} 
       (d g_{\mu\lam}d g_{\nu\s}+ u d g_{\mu\nu}d g_{\lam\s}) ~,  
\ee
where $u >-1/D$ by the positive definiteness of the norm. 

   Before the middle 1980s, pertubation theory had been naively   
defined by replacing the covariant measure with the measure defined 
on the background metric, $\hg$, as $ [g^{-1}dg]_{\hg} [df]_{\hg} $. 
This replacement, however, breaks the background-metric independence, 
or, a infinitesimal version of it, diffeomorphism invariance. 
Renormalizability is also related to the problem of the measure.   
This problem could not be overcome even by going to supergravity. 
It has thus been believed that superstring theory~\cite{gsw} 
is a unique candidate for a background-metric independent quantum theory of 
gravity~\cite{y}.   

  At the end of the 1980s, a new approach emerged.  
This theory was formulated in such a way as to preserve 
the background-metric independence, 
even after the measure is re-expressed in terms of one defined 
based on the background metric. 
Two-dimensional quantum gravity was described exactly 
by Knizhnik, Polyakov and Zamolodchikov (KPZ)~\cite{kpz} 
and David, Distler and Kawai (DDK)~\cite{dk} in this way.    
Subsequently, several 2D models, such as dilaton 
gravity~\cite{h,rt,hks} and  
$R^2$-gravity~\cite{kn}, were studied.    
The idea can apply to even-dimensional quantum gravity, which makes 
the theory background-metric independent. To do this,  
we must add an action, $S$, as in Refs.\cite{hs,h}  
\bb
     Z= \int \frac{[d\phi]_{\hg} [\e^{-h}d \e^h]_{\hg} 
                    [df]_{\bg} }{\hbox{vol(diff.)}} 
                 \exp \bigl[-S(\phi,\bg) - I(f,g) \bigr] ~.
                    \label{zz}
\ee
The metric is now decomposed as $g_{\mu\nu}=\e^{2\phi}\bg_{\mu\nu}$ and 
$\bg_{\mu\nu}=(\hg \e^h)_{\mu\nu}$, where $tr(h)=0$, adopted by 
Kawai, Kitazawa and Ninomiya~\cite{kkn} as well as by the 
present author~\cite{hs}. 
The measures of the metric fields are defined on the background metric 
by the following norms:
\bba
    && <d \phi, d \phi>_{\hg} = \int d^D x \sq{\hg} (d \phi)^2 ~,
            \label{mphi}        \\
    && <d h, d h>_{\hg} 
        = \int d^D x \sq{\hg} ~tr (\e^{-h} d \e^h)^2 ~. 
            \label{mh}
\eea
The action, $S$, is the contribution induced from the measures, and  
it is related to the conformal anomaly. 
Note that the existence of an anomaly usually 
means that a classical symmetry breaks down at the quantum level, 
but now the anomalous contribution, $S$, is needed to ensure  
diffeomorphism invariance at the quantum level.  

   The action $S$ must satisfy the Wess-Zumino condition~\cite{wz} 
to realize  background-metric independence, 
the condition $Z(\e^{2\om}\hg\e^b)=Z(\hg)$, referred to as the BMI condition 
in the following, where $\om$ and $b^{\mu}_{~\nu}$ [with 
$tr(b)=0$] are local functions. Since the measure $[\e^{-h}d\e^h]_{\hg}$, 
itself, is invariant under the change $\e^h \rightarrow \e^{-b}\e^h$, 
the BMI condition for the traceless mode represents the condition 
that $\hg$ and $h$ 
always appear in the combination $\bg=\hg \e^h$ in the theory~\cite{hs}. 
Although this is a rather trivial condition, 
it implies that when we calculate $S$, we can guess the 
$h$ dependence if the $\hg$ dependence of $S$ can be formed.    

  In addition to being background-metric independent, 
namely, invariant under general coordinate transformations, 
\bba
      && \dl \phi = \frac{1}{D} \hnabla_{\lam}\xi^{\lam} +  
                     \xi^{\lam}\pd_{\lam}\phi ~, 
                      \nonumber   \\ 
      && \dl \bg_{\mu\nu} = \bg_{\mu\lam}\bnabla_{\nu}\xi^{\lam} 
                 +\bg_{\nu\lam}\bnabla_{\mu}\xi^{\lam}  
                 -\frac{2}{D}\bg_{\mu\nu}\hnabla_{\lam}\xi^{\lam} ~,    
                      \label{gct}
\eea  
the theory (\ref{zz}) should also be renormalizable. 
In 4 dimensions, $S$ is a fourth order function. In this paper 
we assert that the measure of a 4D gravitational field, 
including $S$ and the fourth order invariant action~\cite{ud}--\cite{bv}, 
are uniquely determined by the two conditions. Thus 
the integrability condition discussed by Riegert, Fradkin and 
Tseytlin~\cite{r}, which is generalized to the form that can be applied 
to higher loops, should be satisfied. 
Here, we emphasize that the fourth order terms are  
required for diffeomorphism invariance, and they are by no means 
introduced by hand only for renormalizability, as in previous 
studies~\cite{dp}--\cite{bv}. 

  This paper is organized as follows. In the next section, 
as an exercise before going to 4D models, we briefly review 
2D quantum gravity from the viewpoint of diffeomorphism invariance. 
The KPZ/DDK formula~\cite{kpz,dk} on a spherical topology is 
computed using dimensional regularization. 
In this model, many fundamental items are included. 
In section 3 we consider the 4D model proposed 
in a previous paper~\cite{hs}, which is slightly modified here.  
In 4 dimensions the traceless mode as well as 
the conformal mode become dynamical. 
Perturbation theory is applied to the traceless mode, 
where a dimensionless coupling constant, $t$, is introduced, 
while the conformal mode can be managed exactly~\cite{kkn,hs}. 
The model studied by Antoniadis, Mazur and Mottola~\cite{am,amm1,amm2} 
as an effective theory of quantum gravity essentially corresponds to 
our model in the limit $t \arr 0$. 
In this section we present evidence that our model 
is renormalizable and diffeomorphism invariant, 
in particular satisfying the 
integrability conditions discussed in~\cite{r} up to order $t^2$, 
or the two-loop order for the vacuum polarization of the traceless mode.   
Higher order renormalizability is also discussed 
from the viewpoint of diffeomorphism invariance.  
At the end of this section we discuss our computations of quantum 
corrections of the cosmological constant up to order $t^2$, 
to which diagrams up to 3-loops contribute.   
The computation is analogous to that for 2D quantum gravity, though the 
equation used to determine the conformal charge of the cosmological 
constant becomes biquadratic at this order.  
It can be solved perturbatively. 
We finally estimate the strength of the coupling $t$ using  
the results of recent numerical simulations~\cite{bbkptt}.
Section 4 is devoted to conclusions and discussion. 
At the end of that section, we discuss proper manner to define 
the $S$-matrix and the method to remove the negative-metric states 
in the fourth order theory required by diffeomorphism invariance. 
Then, the role of the  
$\phi {\bar F}$ term in the Wess-Zumino action is emphasized 
in analogy to  the $R_{\xi}$ gauge in spontaneously broken 
gauge theory.   
     
  Our curvature conventions are $R_{\mu\nu}=R^{\lam}_{~\mu\lam\nu}$ 
and $R^{\lam}_{~\mu\s\nu}=\pd_{\s}\Gamma^{\lam}_{~\mu\nu}-\cdots$. 

\section{Review of A Two-dimensional Model}
\setcounter{equation}{0}
\noindent

 In this section we re-investigate 2D quantum gravity with $N$ massless 
scalar fields, $X$~\cite{kpz,dk}. 
Since we wish to extend these arguments to 4 dimensions,  
we do not use conformal field theories here. 
The Wess-Zumino action $S$ in 2 dimensions, referred to as the Liouville 
action, is given by integrating the 2D conformal anomaly as~\cite{p} 
\bba
    S(\phi,\bg) &=& \fr{a}{4\pi} \int d^2 x \int^{\phi}_0 d \phi \sq{g} R 
                    \nonumber \\ 
                &=& \fr{a}{4\pi} \int d^2 x \sq{\hg} \bigl( 
                          \bg^{\mu\nu}\pd_{\mu} \phi \pd_{\nu} \phi 
                                                    + \bR \phi \bigr) ~. 
\eea
Here, the Wess-Zumino condition is defined by 
\bb
       S(\phi,\bg)= S(\om,\bg) + S(\phi-\om,\e^{2\om}\bg) ~. 
           \label{wz}
\ee
One can easily prove that $S$ satisfies the above equation by dividing 
the region of integration $[0,\phi]$ into $[0,\om]$ and $[\om,\phi]$.

   The coefficient $a$ is determined by requiring diffeomorphism invariance.  
Let us apply the general coordinate transformation (\ref{gct}) and 
$\dl X = \xi^{\lam}\pd_{\lam}X$ to expression (\ref{zz}). 
The action $I$ is invariant, so that $\dl I(X,g)=0$, 
but $S$ is not. We obtain the following expression: 
\bb
    \dl S(\phi,\bg)
      =\frac{a}{8\pi} \int d^2 x \sq{\hg} \hnabla_{\lam}\xi^{\lam}\bR ~. 
               \label{iwz}
\ee
This is easily proved by decomposing the general coordinate transformation 
into two parts, $\dl=\dl_c + \dl_{\om}$, and noting that 
$S$ is invariant under $\dl_c$, 
which is a transformation law in which 
$\phi$ transforms as if it were a scalar field 
on the metric $\bg_{\mu\nu}$ such that 
$\dl_c \bg_{\mu\nu} = \bg_{\mu\lam}\bnabla_{\nu}\xi^{\lam} 
+\bg_{\nu\lam}\bnabla_{\mu}\xi^{\lam}$ 
and $\dl_c\phi = \xi^{\lam}\pd_{\lam}\phi$, 
while $\dl_{\om}$ is the conformal change of the metric $\bg_{\mu\nu}$ 
and the local shift of $\phi$ defined by  
$\dl_{\om} \bg_{\mu\nu} = 2\om \bg_{\mu\nu}$ 
and $\dl_{\om} \phi = -\om$, with  
\bb
      \om=-\half \hnabla_{\lam}\xi^{\lam} ~. 
               \label{om2d}
\ee
Since $\xi$ is infinitesimal, the r.h.s. of (\ref{iwz}) can be 
expressed as $-S(\om, \bg )$ using (\ref{om2d}). 
Thus, (\ref{iwz}) is identically the Wess-Zumino condition (\ref{wz}) 
with an infinitesimal value (\ref{om2d}).

 On the other hand, the measure itself is not invariant, 
due to the fact that there is no $\e^{2\phi}$ factor in the norm, 
which conformally changes in the form  \\ %%%%%%%%(break)%%%%% 
$[d\phi]_{(1+2\om)\hg} [\e^{-h}d \e^h]_{(1+2\om)\hg} [dX]_{(1+2\om)\hg}$, 
where $\om$ is given by (\ref{om2d}). This is easily seen by applying 
the general coordinate transformation (\ref{gct}) to the norms defined 
on the background metric as, for instance, 
\bba
   && <d\phi^{\pp}, d\phi^{\pp}>_{\hg}  
      = \int d^2 x \sq{\hg}(d\phi^{\pp})^2 
              \nonumber  \\ 
   && = \int d^2 x \sq{\hg}(1-\hnabla_{\lam}\xi^{\lam})(d\phi)^2
      = <d\phi , d\phi >_{(1+2\om)\hg} ~,  
            \label{ngct}
\eea 
where $\phi^{\pp}=\phi+\dl\phi$, so that 
$d\phi^{\pp}=d\phi + \xi^{\lam} \pd_{\lam} d\phi$. 
In the second equality of (\ref{ngct}), 
the $o(\xi^2)$ term has been omitted.   
 
 We first evaluate the measure of the traceless mode. Let us take the 
gauge-fixing condition  
$({\hat P}_1^{\dagger}\bg)_{\mu}=\hnabla^{\lam}\bg_{\lam\mu}=0$ 
to be  as usual, where $P_1$ is the projection operator to the 
traceless part such that $\dl\bg_{\mu\nu}=({\bar P}_1\xi)_{\mu\nu}$. 
We then obtain  
\bb
      \frac{[\e^{-h}d\e^h]_{\hg}} {\hbox{vol(diff.)}}
       = [\e^{-h}d\e^h]_{\hg} \det {\hat P}_1^{\dagger}{\bar P}_1 
           \dl ({\hat P}_1^{\dagger}\bg ) ~. 
\ee
As is wellknown, in two dimensions, the r.h.s. can be further 
written in the form~\cite{p} 
\bb
     [d\tau] (\det {\bar P}_1^{\dagger}{\bar P}_1 )^{\half} ~, 
\ee        
where $[d\tau] = [\e^{-h}d\e^h]_{\hg} \det {\hat P}_1^{\dagger} 
\dl ({\hat P}_1^{\dagger}\bg ) $ 
represents integration over the Teichm\"{u}ller parameter. 
Note that this rewriting is equivalent to taking the gauge condition 
as $h^{\mu}_{~\nu}= 0$~\footnote{%%%%%%%%%%%(footnote 3)%%%%%%%%%%%%
If instead one chooses the ghost determinant as 
$\det {\hat P}_1^{\dagger}{\bar P}_1$, one must treat the traceless 
mode as a quantum field~\cite{w,kkn}. 
}  %%%%%%%%%%(end of footnote 3)%%%%%%%%%%
up to the zero mode, and hence obtain a gauge-fixing term   
$i B_{\mu\nu} h^{\mu\nu}$,    
where $B_{\mu\nu}$ is a symmetric traceless field. 
The ghost action is then given by 
$b_{\mu\nu}\dl h^{\mu\nu} \vert_{\xi^{\lam}=c^{\lam}}$, 
which is the well-known $bc$-ghost system,  
where $b_{\mu\nu}$ is the anti-ghost related to $B_{\mu\nu}$ and $c^{\lam}$ 
is the reparametrization ghost. 
Using the well-known conformal property of the $P_1$ operator~\cite{p}, 
we obtain 
\bb
      [\e^{-h}d\e^h]_{(1+2\om)\hg} = [\e^{-h}d\e^h]_{\hg} 
               \biggl( 1+\frac{26}{48\pi} 
                 \int d^2 x \sq{\hg} \hnabla_{\lam}\xi^{\lam}\bR  
                           \biggr) ~.
                   \label{cp-h}
\ee

   The measure of $\phi$ is easily evaluated by using the bilinear  
term of the action, ${\cal I}= S+I$, as a regulator, which is that 
of a scalar field. Combined with the results for the traceless mode 
(\ref{cp-h}) and $N+1$ scalar fields,  we obtain
\bba
    && [dX]_{(1+2\om)\hg}[d\phi]_{(1+2\om)\hg}     
       [\e^{-h}d\e^h]_{(1+2\om)\hg}\exp (-{\cal I}) 
            \nonumber \\ 
    && \quad = (1-U(\om)) [dX]_{\hg}[d\phi]_{\hg} 
             [\e^{-h}d\e^h]_{\hg}\exp (-{\cal I}) ~,
\eea
where
\bb
     U(\om)  = - \frac{25-N}{48\pi} \int d^2 x 
               \sq{\hg} \hnabla_{\lam}\xi^{\lam}\bR ~. 
                  \label{ca2}
\ee
Thus, the variation of the partition function (\ref{zz}) w.r.t. the 
general coordinate transformation becomes
\bb
    \dl Z = \fr{-6a+25-N}{48\pi}  
                \biggl<  \int d^2 x 
               \sq{\hg} \hnabla_{\lam}\xi^{\lam}\bR \biggr> ~.
\ee
Therefore, the diffeomorphism invariance requires~\cite{dk}   
\bb
        a = \fr{1}{6}(25-N) ~.  \label{va}
\ee       
This is equivalent to the condition that the algebra of 
the energy-momentum tensor, namely, the Virasoro algebra is closed 
without central charge at the quantum level.  

  The lower-dimensional operator generally receives a non-trivial 
quantum correction. The cosmological constant operator, 
$\Lam \int \e^{\a \phi}$, for instance, where classically 
$\a=\hbox{dim} [\Lam]=2$, 
recieves the quantum correction 
\bb
     \a = \hbox{dim} [\Lam] = 2+\gm_{\Lam} ~,
\ee 
where $\gm_{\Lam}$ is the anomalous dimension. In the following we 
consider the perturbation theory on $\a$ in a flat background as 
$\e^{\a\phi}=\sum \frac{\a^n}{n!} \phi^n$, and we calculate $\gm_{\Lam}$ 
using dimensional regularization~\cite{am,amm2}.  

  The combined action in $D=2-2\eps$ dimensions is 
\bb
     {\cal I} = \frac{1}{4\pi}\int d^D x \sq{\hg}\biggl[ 
             a  \Bigl( \bg^{\mu\nu} \pd_{\mu}\phi \pd_{\nu}\phi 
             + {\bar R}\phi \Bigr) + \Lam \e^{\a\phi}  
             + \bg^{\mu\nu} \pd_{\mu}X \pd_{\nu}X  \biggr]  ~,
\ee  
where the $bc$-ghost is omitted. 
The conformal mode $\phi$ and the constant $\a$ are now dimensionless, 
while the field $X$ is of dimension $\mu^{-\eps}$, where $\mu$ is an  
arbitrary mass scale. 
The strategy to compute the anomalous dimension using  dimensional 
regularization is as follows. Consider the coefficient $a$ in front 
of the Wess-Zumino action to be a bare coupling 
of dimension $\mu^{-2\eps}$. 
We carry out a renormalization procedure with $a$ being unknown, 
and compute the anomalous dimension. 
After regularization we take $\eps \arr 0$ and substitute the value 
of $a$ required by diffeomorphism invariance into 
regularized quantities.  

  Let us evaluate loop diagrams in the flat background. 
We can easily see that simple poles come only from the one-loop 
diagram in Fig.~1.
The sum of the two-loop diagram (a) and the associated diagram (b), 
with the one-loop counterterm insertion shown in Fig.~2, yields only 
double poles proportional to $\a^4$.  
The higher-loop diagrams also do not yield simple poles.
%%%%%%%%%%%%%%(diagrams Figs. 1, 2)%%%%%%%%
%One-loop corrections of cos. const. (2 dim.)
\begin{center}
\begin{picture}(100,100)(0,-10)
\CArc(50,50)(20,0,360)
\Vertex(50,30){1}
\Text(52,12)[]{$\cdots$}
\Line(38,10)(50,30)
\Line(42,9)(50,30)
\Line(62,10)(50,30)
\end{picture} \\
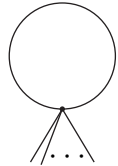
Fig. 1: One-loop contribution to the cosmological constant.
\end{center}
%Two-loop corrections of cos. const. (2 dim.)
\begin{center}
\begin{picture}(300,100)(0,-10)
\Oval(83,50)(22,12)(40)
\Oval(117,50)(22,12)(-40)
\Vertex(100,38){1}
\Text(102,12)[]{$\cdots$}
\Line(88,10)(100,38)
\Line(92,9)(100,38)
\Line(112,10)(100,38)
\Text(100,-3)[]{(a)}

\CArc(200,50)(20,0,360)
\GCirc(200,30){4}{1}
\Line(197,33)(203,27)
\Line(197,27)(203,33)
\Text(202,12)[]{$\cdots$}
\Line(188,10)(200,26)
\Line(192,9)(200,26)
\Line(212,10)(200,26)
\Text(200,-3)[]{(b)}
\end{picture} \\ 
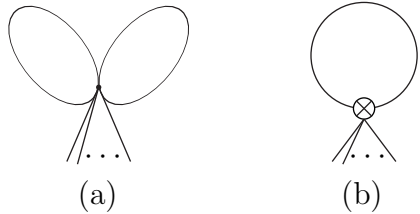
Fig. 2: Two-loop contributions to the cosmological 
constant. \\
The contribution of a one-loop counterterm is also depicted.
\end{center} 
%%%%%%%%%%%%%%%%%%%%%%%%(end of Fig. 1, 2)%%%%%%

  Hence, the renormalization factor defined by the relation 
$\Lam = Z_{\Lam} \Lam_r$ is computed as  
\bb
     Z_{\Lam} = 1-\frac{\a^2}{4\ta}\frac{1}{\eps} 
                 + \frac{\a^4}{32\ta^2} \frac{1}{\eps^2} + \cdots~, 
\ee 
where $\ta$ is the dimensionless bare coupling 
defined by $a=\ta\mu^{-2\eps}$. 
Since the bare coupling, $a$, should be independent of the arbitrary 
mass scale, $\mu$, we obtain a finite value in the limit $\eps \arr 0$:  
\bb
    \gm_{\Lam} = \frac{\mu}{Z_{\Lam}} \frac{d Z_{\Lam}}{d\mu} 
           = \frac{\a^2}{2 a} ~. 
\ee
This value is exact to all orders of the perturbation. 
We thus obtain the quadratic equation $\a=2+\frac{\a^2}{2a}$. 
Solving this equation, we obtain     
\bb
      \a = a - \sq{a^2-4a} ~,   \label{kpz}
\ee
where we choose a solution such that $\a \rightarrow 2$ in the 
classical limit $a \rightarrow \infty$.  

   We now re-explain diffeomorphism invariance in the context of 
dimensional regularization. 
In the curved background we obtain the one-loop exact counterterm  
\bb
    \Delta{\cal I} = - \frac{25-N}{48\pi}\frac{\mu^{-2\eps}}{\eps} 
             \int d^D x \sq{\hg}\bR  ~. 
               \label{ctr}
\ee
{}From Duff's argument~\cite{duff}, 
the variation of this counterterm w.r.t. the general 
coordinate transformation, which is now converted into the conformal 
change of the background metric, 
$ \dl_{\om}\hg_{\mu\nu}=2\om \hg_{\mu\nu}$, 
gives the conformal anomaly $U(\om)$ (\ref{ca2}),  
because the conformal change in the  
scalar curvature in $D=2-2\eps$ dimensions is given by the identity
\bb
      \dl_{\om} \sq{\hg} \bR = -\eps 2\om \sq{\hg} \bR  
                       = \eps  \hnabla_{\lam}\xi^{\lam}\sq{\hg}\bR ~.
\ee 
We thus obtain the value (\ref{va}), as explained above. 
Substituting this value into expression (\ref{kpz}), 
we obtain the well-known KPZ/DDK formula~\cite{kpz,dk} 
for a spherical topology.

\section{Four Dimensional Model} 
\setcounter{equation}{0}
\noindent

  In this section we consider a slightly modified version of a model 
proposed in Ref.~\cite{hs}. 
In 4 dimensions the Wess-Zumino action becomes fourth order, 
parametrized by the three constants $a$, $b$ and $c$ in the form
\bba
    S(\phi,\bg) &=& \frac{1}{(4\pi)^2} \int d^4 x \int^{\phi}_0 
            d \phi  \sq{g}  \biggl[ 
            a \biggl( F+\frac{2}{3}\Box R \biggr) 
            + b G + c \Box R  \biggr]  
                 \nonumber \\ 
        &=& \frac{1}{(4\pi)^2} \int d^4x \biggl[ \sq{\hg} 
            \biggl\{  a {\bar F} \phi 
            +2b \phi {\bar \Delta}_4 \phi 
            +b \Bigl( {\bar G}-\fr{2}{3} 
            \bBox {\bar R} \Bigr) \phi \biggr\}            
                  \nonumber \\ 
        && \qquad\qquad\qquad
          - \frac{1}{36}(2a+2b+3c) 
          \Bigl( \sq{g} R^2 
          -\sq{\hg}{\bar R}^2 \Bigr) \biggr] ~,   
\eea
where the addition of $c$ is a modification. 
$F$ is the square of the Weyl tensor, and $G$ is the Euler density.   
They are defined by 
\bba
    F &=& R_{\mu\nu\lam\s}R^{\mu\nu\lam\s}-2R_{\mu\nu}R^{\mu\nu} 
           +\fr{1}{3}R^2 ~,
               \\ 
    G &=& R_{\mu\nu\lam\s}R^{\mu\nu\lam\s}-4R_{\mu\nu}R^{\mu\nu} 
           +R^2 ~. 
\eea 
The operator $\Delta_4$ is the conformally covariant fourth 
order operator~\cite{r} defined by  
\bb
      \Delta_4 = \Box^2 
               + 2 R^{\mu\nu}\nabla_{\mu}\nabla_{\nu} 
                -\fr{2}{3}R \Box 
                + \fr{1}{3}(\nabla^{\mu}R)\nabla_{\mu}  
\ee
and satisfying $\Delta_4 = \e^{-4\phi}{\bar \Delta}_4$. 
Thus, {\it in order to ensure diffeomorphism invariance 
in 4 dimensions, the theory must be fourth order 
in the gravity sector}. Only in that case is 
the theory renormalizable also.  

  Here, note that $R^2$ is not integrable w.r.t. the conformal mode, 
which is the reason that the number of the independent parameters is three. 
This fact is related to the integrability condition~\cite{r} 
discussed below. 
 
  Let us apply the general coordinate transformation (\ref{gct}) 
to the theory (\ref{zz}) in 4 dimensions. The Wess-Zumino action 
then changes by 
\bb
     \dl S(\phi, \bg) = \fr{1}{4(4\pi)^2} \int d^4 x
                     \sq{\hg} \hnabla_{\lam}\xi^{\lam} 
                \biggl[ a \biggl( {\bar F} + \fr{2}{3}\bBox \bR 
                           \biggr) +b {\bar G} 
                           + c \bBox \bR \biggr] ~. 
\ee       
As in two dimensions, this is simply the Wess-Zumino condition 
(\ref{wz}) for the infinitesimal value  
\bb
      \om=-\fr{1}{4} \hnabla_{\lam}\xi^{\lam} ~. 
                \label{om4d}
\ee 
The measure is also not invariant under the transformation (\ref{gct}), 
which conformally changes in the form 
$[d\phi]_{(1+2\om)\hg} [\e^{-h}d \e^h]_{(1+2\om)\hg} [df]_{(1+2\om)\bg}$ 
with (\ref{om4d}).
The coefficients $a$, $b$ and $c$ are determined by imposing the condition 
of diffeomorphism invariance, so that the $\xi$ dependences from 
the action and the measures cancel out. 

  The equations to determine the coefficients $a$, $b$ and $c$ are now 
not so simple.  Consider the case that the $\xi$ dependence of 
the deformed measure can be re-expressed in the form  
$(1-U(\om)) [d\phi]_{\hg} [\e^{-h}d \e^h]_{\hg}[df]_{\bg}$ 
using the expression of the conformal anomaly   
\bb
     U(\om) = \fr{1}{(4\pi)^2} \int d^4 x  \sq{\hg} ~\om  
            \biggl[ a^{\pp} \biggl( {\bar F} + \fr{2}{3}\bBox \bR 
                   \biggr) + b^{\pp} {\bar G} 
                   + c^{\pp} \bBox \bR \biggr] ~.   
\ee  
The regulator used to evaluate the measures is now the combined 
action ${\cal I}= S+I$, which includes the unknown constants $a$, $b$ 
and $c$, so that the coefficients $a^{\pp}$, $b^{\pp}$ and $c^{\pp}$ 
become in general functions of $a$, $b$ and $c$. 
Thus, the BMI condition, $U+\dl S=0$, is now expressed by  
the equations 
\bb
      a^{\pp}(a,b,c) =a ~, \qquad  b^{\pp}(a,b,c)=b ~, 
      \qquad c^{\pp}(a,b,c)=c ~. 
             \label{bmi}
\ee      

  To be able to impose this BMI condition it is necessary that  
the $\xi$ dependence of the measure can indeed be expressed in 
the above form.  
This forces the matter fields to be coupled with gravity conformally.  
As for the gravity sector, we must consider the 
fourth order invariant action, whose form is restricted by the 
conditions of renormalizability and diffeomorphism invariance 
discussed below. 
 
  We consider the following invariant action, including 
fourth order terms:~\footnote{%%%%(footnote 4)%%%%%%
Note that the coefficient $d$ corresponds to $c$ in Ref.\cite{hs}. 
} %%%%%%%%%%%(end of footnote 4)%%%%%%
\bb
     I(A,g)=I_{\rm 4th} +I_{LE} ~,  \label{cla}
\ee
where  
\bba
     && I_{\rm 4th} = \fr{1}{(4\pi)^2} \int d^4 x \sq{g} 
               \biggl( \fr{1}{t^2}  F +  d  R^2 \biggr) ~, 
                      \\ 
     &&  I_{LE} = \fr{1}{(4\pi)^2} \int d^4 x \sq{g} 
                   (- m^2 R + \Lambda ) + I_A (A,g) ~.    
\eea    
Here, $I_{\rm 4th}$ is the fourth order part, which could be regarded as 
being a part of the measure. $I_{LE}$ is the usual second order 
action, which describes low-energy physics. 
$m^2$ is the inverse of the gravitational constant,  
and $\Lambda$ is the cosmological constant. 
In this section we consider $U(1)$ gauge fields as an example. 
$I_A$ is the action of the $N_A$ gauge fields.  

  Above, we introduced the dimensionless self-coupling constant 
$t$ only for the traceless mode in the form~\cite{kkn,hs}
\bb
          \bg_{\mu\nu}=(\hg \e^{th})_{\mu\nu} 
                = \hg_{\mu\lam} \biggl( \dl^{\lam}_{~\nu} 
                   + t h^{\lam}_{~\nu}  
                   + \frac{t^2}{2} (h^2)^{\lam}_{~\nu} + \cdots 
                   \biggr) ~.              
\ee
The general coordinate transformation for the traceless mode is then 
expressed as 
\bba
   t\dl h^{\mu}_{~\nu} &=& \hnabla^{\mu} \xi_{\nu} 
                       +\hnabla_{\nu} \xi^{\mu}
           - \half \dl^{\mu}_{~\nu}  
                    \hnabla_{\lam} \xi^{\lam}   
                 + t \xi^{\lam} \hnabla_{\lam} 
                                 h^{\mu}_{~\nu}   
                \nonumber    \\ 
       &&
           + \fr{t}{2} h^{\mu}_{~\lam} 
                \Bigl( \hnabla_{\nu} \xi^{\lam} 
                  - \hnabla^{\lam} \xi_{\nu} \Bigr) 
            + \fr{t}{2} h^{\lam}_{~\nu} 
                \Bigl( \hnabla^{\mu} \xi_{\lam} 
                  - \hnabla_{\lam} \xi^{\mu} \Bigr) 
             + \cdots ~,
                   \label{hgct} 
\eea
where $\xi_{\mu}= \hg_{\mu\lam}\xi^{\lam}$. 
The BMI condition (\ref{bmi}) can then be solved using a  
perturbation in $t$. 

  In the following we consider a model 
with~\cite{hs}
\bb
      d=\fr{1}{36}(2a+2b+3c) ~.  \label{d}
\ee
This condition implies that the self-interaction terms of 
the conformal mode, namely $R^2$ terms in the combined action, 
cancel out. 
This model is analogous to the 2D quantum gravity discussed 
in section 2. 
In this paper we assert that {\it only in this case does the theory 
become renormalizable and diffeomorphism invariant 
simultaneously in the perturbation of the traceless mode, 
where the coefficients $a$, $b$ and $c$ are uniquely determined 
by the diffeomorphism invariance}. 

\subsection{Renormalization}
\noindent

  To confirm the above assertion, we investigate the renormalizability 
and diffeomorphism invariance of our model focusing on how crucial   
condition (\ref{d}) is. 
We use dimensional regularization to regularize our model 
as a theory on the metric $\bg_{\mu\nu}$.  
Here, we assume the renormalizability of our model  
and derive general formulae 
for renormalization and diffeomorphism invariance. 
In the following subsections, we show that no inconsistency 
appears in our model, at least up to order $t^2$.   

 In this paper we consider the $m^2=0$ case for simplicity. 
Thus, the bare combined action becomes
\bba
   &&{\cal I} = \frac{1}{(4\pi)^2} \int d^D x \sq{\hg}
        \biggl[  2 b \phi  {\bar \Delta}_4 \phi  + a {\bar F} \phi
         + b \Bigl( {\bar G}-\fr{2}{3} \bBox {\bar R} \Bigr) \phi 
              \nonumber   \\
   && \qquad\qquad\qquad
        +\frac{1}{t^2}{\bar F} + \frac{1}{36}(2a+2b+3c){\bar R}^2 
        + \Lam \e^{\a\phi}   \biggr] + I_A(A,\bg) ~,
                \label{act}
\eea 
where $D=4-2\eps$. The constants $a$, $b$ and $c$ are considered 
to be bare couplings of dimension $\mu^{-2\eps}$, and $t$ is 
the bare coupling of dimension $\mu^{\eps}$. 
For later use we introduce the dimensionless bare quantities 
$\ta$, $\tb$, $\tc$ and $\tt$, defined by the relations 
$a=\ta \mu^{-2\eps}$, $b=\tb \mu^{-2\eps}$, $c=\tc \mu^{-2\eps}$ 
and $t = \tt \mu^{\eps}$.  
Here, we set the dimension of the conformal mode to zero, 
while the traceless mode is of dimension $\mu^{-\eps}$.  
 
 The renormalizations of the quantum fields and couplings are defined by 
the following relations:
\bb
     \phi = Z^{1/2}_{\phi} \phi_r ~, \qquad  
     h^{\mu}_{~\nu}= Z^{1/2}_h h^{\mu}_{r \nu}  
\ee
and
\bb
     a=Z_a a_r ~, \qquad   b=Z_b b_r ~, \qquad  c=Z_c c_r ~, \qquad
     t=Z_t t_r ~. 
\ee
Since the quantum traceless mode must be gauge-fixed to carry out  
calculations, we introduce here a more well-behaved variable, 
the background field~\cite{ab} for the traceless mode defined by 
\bb
      \hg_{\mu\nu}=(\e^{t\hh})_{\mu\nu} ~,  
             \label{bm}
\ee
where $\hh^{\mu}_{~\mu}=0$.  
The renormalization of the background traceless mode is also defined 
by the relation~\cite{ab} 
\bb
     \hh^{\mu}_{~\nu}= Z^{1/2}_{\hh} \hh^{~\mu}_{r ~\nu} ~.        
\ee 

  On the other hand, since the dynamics of gravitational fields are   
governed by fourth order derivative theory, and  matter fields are coupled 
to gravitational fields conformally, the counterterms for the background 
field become fourth order:   
\bba
  && \Delta {\cal I}_F 
     = - P_F \frac{\mu^{-2\eps}}{(4\pi)^2}\int d^D x 
         \sq{\hg_r}{\hat F}_r ~, \quad  
             \\ 
  && \Delta {\cal I}_G 
     = - P_G \frac{\mu^{-2\eps}}{(4\pi)^2}\int d^D x 
          \sq{\hg_r}{\hat G}_r ~, 
             \\ 
  && \Delta {\cal I}_{R^2} 
     = - P_{R^2}\frac{\mu^{-2\eps}}{(4\pi)^2}\int d^D x 
         \sq{\hg_r}{\hat R}_r^2 ~.
\eea
Here ${\hat F}_r$, ${\hat G}_r$ and ${\hat R}_r^2$ are the square of  
the Weyl tensor, the Euler density, and the square of the scalar curvature 
written in terms of renormalized quantities, respectively.  
The pole terms, $P_F$, $P_G$ and $P_{R^2}$, are expanded as 
\bba
    && P_F = \frac{f_0}{\eps} 
                        + \tt_r^2 \biggl(\frac{f_1}{\eps^2} 
                        +\frac{f_1^{\pp}}{\eps} \biggr) 
                       + \cdots   ~,
                \label{pt1}         \\
    && P_G =  \frac{g_0}{\eps} + \tt_r^2 \biggl(\frac{g_1}{\eps^2} 
                        +\frac{g_1^{\pp}}{\eps} \biggr)       
                         + \cdots ~,
                           \\  
    && P_{R^2} = \frac{l_0}{\eps} + \tt_r^2 \biggl(\frac{l_1}{\eps^2} 
                        +\frac{l_1^{\pp}}{\eps} \biggr) 
                        + \cdots ~,
                        \label{pt3}
\eea  
where $\tt_r$ is the dimensionless renormalized coupling. 
The coefficients $f_n$, $g_n$, $l_n$, and so on, are generally functions 
of the dimensionless renormalized couplings, $\ta_r$, $\tb_r$ and $\tc_r$.  

  The counterterms, $\Delta {\cal I}_F$ and $\Delta {\cal I}_G$, are 
conformally invariant at $D=4$. 
In particular, $\Delta {\cal I}_G$ becomes topological, 
which is just an analog of the counterterm (\ref{ctr}) 
in 2 dimensions. In $D=4-2\eps$ dimensions, the square of the Weyl tensor 
and the Euler density satisfy the following identities~\cite{duff}:
\bba
   && \dl_{\om} \sq{\hg_r}{\hat F}_r 
         = -2 \eps \om \sq{\hg_r} \biggl( {\hat F}_r 
                     +\frac{2}{3}\hBox \hR_r \biggr) ~, 
                      \\ 
   && \dl_{\om} \sq{\hg_r}{\hat G}_r 
         = -2 \eps \om \sq{\hg_r}{\hat G}_r  ~.                    
\eea
On the other hand, the counterterm $\Delta {\cal I}_{R^2}$ is not 
conformally invariant at $D=4$, because $\hR^2$ satisfies the identity 
\bb
    \dl_{\om} \sq{\hg_r}\hR^2 = -12 \om \sq{\hg_r}\hBox \hR ~. 
               \label{vosr}
\ee
This counterterm is related to the renormalizations of the 
couplings $a$, $b$ and $c$, and also the integrability condition of 
the Wess-Zumino action discussed in~\cite{r}, which is discussed 
in detail in section 3.3. 
    
  The renormalization of the background traceless mode is now defined 
by the relation
\bb
     \frac{1}{\tt^2} {\hat F} = \frac{1}{\tt_r^2}{\hat F}_r 
                 - P_F {\hat F}_r ~. 
\ee 
We thus obtain the renormalization factor of the coupling,
\bb
      (Z_t)^{-2} = 1 -  P_F \tt_r^2  ~,
\ee 
and the relation 
\bb
              Z_t Z_{\hh}^{1/2} = 1 ~. \label{zth}
\ee
Therefore, the background metric is not renormalized, and hence   
$\hg_{r\mu\nu} = \hg_{\mu\nu}$.

\subsection{Some results on renormalization factors}
\noindent

  We present here some results on the renormalization factors 
discussed in the previous subsection.  
Let us first compute the coefficients $f_0$, $g_0$ and $l_0$ in 
the counterterms. They have already been calculated in 
Refs.\cite{amm1,hs} by considering the case $t \arr 0$ 
in a curved background.\footnote{ %%%%%%%%(footnote 5)%%%%
In this calculation we do not specify the background metric 
to be in the form (\ref{bm}).
} %%%%%%%%%%(end of footnote 5)%%%%%
The interaction terms among quantum fields then vanish, 
and the kinetic terms reduce to conformal invariant 
ones due to condition (\ref{d}), so that 
\bb
      l_0=0 ~.  \label{l0v}
\ee
The coefficients $f_0$ and $g_0$ are given by the values   
\bb
      f_0 = -\fr{N_A}{20}-\fr{197}{60} ~, \qquad
      g_0 = \fr{31 N_A}{360}+\fr{769}{360} ~.            
              \label{fg}
\ee
{}From this we obtain $Z_t = 1+  \frac{f_0 \tt^2_r}{2\eps} + O(\tt_r^4)$,  
and hence the $\beta$-function  
\bb
      \beta = -\eps \tt_r - \tt_r \frac{\mu}{Z_t}\frac{d Z_t}{d\mu} 
            = f_0  t_r^3 + \cdots ~. 
\ee 
Since $f_0$ is negative, this model is asymptotically free~\cite{ft1,t}. 

  Next, we see that the Wess-Zumino action is likely to be 
unrenormalized to order $t^2_r$. 
First, consider the self-energy diagrams of the conformal mode in a  
flat background. The combined action is expanded as 
\bb
     {\cal I} = \frac{1}{(4\pi)^2} \int d^D x  
           \Bigl( {\cal L}_{\rm kin} + {\cal L}_{\rm int}^{(2)} 
      + {\cal L}_{\rm int}^{(3)} + {\cal L}_{\rm int}^{(4)} + \cdots 
            \Bigr) ~. 
\ee
The kinetic terms for the quantum gravitational fields 
are~\footnote{%%%%%%%%(footnote 6)%%%%%%%
Here, we use the relation 
${\bar F}={\bar G}+ 2 (\bR^{\mu\nu}\bR_{\mu\nu} - \frac{1}{3} \bR^2 )$. 
The second combination of r.h.s. gives the kinetic term of 
the traceless mode.   
} %%%%%%%%%%%(end of footnote 6)%%%%%%%%
\bb
     {\cal L}_{\rm kin} = 2 b ~\phi \Box^2 \phi 
           + \half h_{\mu\nu} \Box^2 h^{\mu\nu} 
         - \pd^{\mu}\chi^{\nu} \pd_{\mu} \chi_{\nu} 
         + \biggl(\frac{1}{3} +d t^2 \biggr) 
           \Bigl( \pd_{\lam}\chi^{\lam} \Bigr)^2 ~,
                  \label{kin}
\ee
where $\chi^{\mu}=\pd^{\lam}h^{\mu}_{~\lam}$, and 
$d$ is given by (\ref{d}). In this subsection, we take 
$\Box =\pd^{\lam}\pd_{\lam}$ for simplicity.  
The interaction terms used here are 
\bba
    && {\cal L}_{\rm int}^{(2)} = -\frac{2}{3} b t 
                \Box \phi \pd_{\mu}\pd_{\nu} h^{\mu\nu} ~,
                    \\ 
    && {\cal L}_{\rm int}^{(3)} = 2 b t \biggl( 
          2\pd_{\mu} \phi \pd_{\nu} \Box \phi 
          + \frac{4}{3}\pd_{\mu}\pd_{\lam}\phi \pd_{\nu}\pd^{\lam}\phi 
                  \nonumber \\ 
    && \qquad\qquad\qquad 
        -\frac{2}{3}\pd^{\lam}\phi \pd_{\mu}\pd_{\nu}\pd_{\lam} \phi 
          -2\pd_{\mu}\pd_{\nu}\phi \Box \phi 
             \biggr)~ h^{\mu\nu} ~,                
                    \\  
    && {\cal L}_{\rm int}^{(4)} = 2 b t^2 \Bigl( 
         \Box\phi\pd_{\mu}\pd_{\nu} \phi h^{\mu}_{~\lam} h^{\lam\nu} 
         +\pd_{\mu}\pd_{\nu}\phi \pd_{\lam}\pd_{\s}\phi 
               h^{\mu\nu}h^{\lam\s} 
               \nonumber \\ 
    && \qquad\qquad\qquad\qquad 
         + ~\hbox{terms including $\pd h$} ~\Bigr)~, \label{vx4}         
\eea
where ${\cal L}^{(2)}_{\rm int}$ is derived  from the term 
$-\frac{2}{3}b \bBox\bR\phi$. The vertices ${\cal L}_{\rm int}^{(3)}$ 
and ${\cal L}_{\rm int}^{(4)}$ are from 
$2b \phi {\bar \Delta}_4 \phi$. 
These vertices are, of course, decomposed into the renormalized ones 
and their counterterms.     

  The propagator of the conformal mode, which is described by a line,  
is given by 
\bb
    \frac{(4\pi)^2}{4\tb_r} \frac{\mu^{2\eps}}{k^4} ~.  
\ee
For the traceless mode  we consider the gauge-fixing term  
\bb
      {\cal L}_{GF} = -\frac{1}{\zeta} 
         \chi^{\mu} N_{\mu\nu} \chi^{\nu}  ~, 
       \label{gfix}
\ee
where 
\bb
      N_{\mu\nu}= \Box\dl_{\mu\nu} -\biggl( \frac{1}{3}+dt^2 \biggr) 
                   \pd_{\mu}\pd_{\nu} ~. 
\ee 
Here, we take the Feynman-type gauge $\zeta= 1$   
in which the gauge-fixing term cancels out the last two 
of the kinetic terms in (\ref{kin}). 
Then, the propagator of the traceless mode, which is described by 
the wavy line, becomes 
\bb
        (4\pi)^2 \frac{1}{k^4} (I_H)^{\mu\nu}_{~~, \lam\s} ~, 
\ee
where $I_H$ is the projection operator to the traceless mode, defined by
\bb
     (I_H)^{\mu\nu}_{~~, \lam\s} 
        = \dl^{\mu}_{(\lam}\dl^{\nu}_{\s)} 
             - \frac{1}{D}\dl^{\mu\nu}\dl_{\lam\s} ~,  
\ee
which satisfies $I_H^2 =I_H$. 
The ghost action is given by applying the general coordinate 
transformation to the gauge-fixing condition, 
which is not coupled with the conformal mode.  

  Diagrams (a) and (b) in Fig.~3 make contributions of order  
$t_r^2$ to $Z_b Z_{\phi}$, while the diagram in Fig.~4, for example,  
does not contribute, because such a diagram now 
becomes order $t_r^4$, where the vertices in Fig. 4 come from the terms  
linear in $\phi$. 
In general, diagrams which make contributions proportional to 
$\ta_r^k \tb_r^l \tc_r^m \tt_r^{2n}$ to renormalization factors  
have $n-k-l-m$ loops. 
%%%%%%%%%%%%%%(diagrams Figs.3,4)%%%%%%% 
%Self-energy diagarms a,b 
\begin{center}
\begin{picture}(430,60)(0,-10)
\Line(100,10)(200,10)
\PhotonArc(150,10)(30,0,180){3}{8.5} 
\Vertex(120,10){1}
\Vertex(180,10){1}
\Text(150,-3)[]{(a)}    

\Line(250,10)(330,10)
\PhotonArc(290,31)(18,22,382){3}{12}
\Vertex(290,10){1}
\Text(290,-3)[]{(b)} 
\end{picture} \\ 
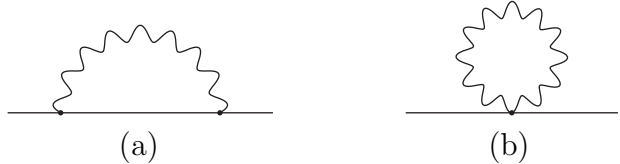
Fig. 3: One-loop contributions of order $t_r^2$ to $Z_{\phi}$.
\end{center} 
%Self-energy diagram c
\begin{center}
\begin{picture}(300,60)(0,0)
\PhotonArc(150,30)(20,-6,354){3}{16} 
\Line(100,30)(127,30)
\Line(173,30)(200,30)
\Vertex(127,30){1}
\Vertex(173,30){1}
\end{picture} \\ 
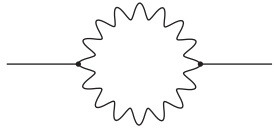
Fig. 4: One-loop, but the $t_r^4$-order contribution to $Z_{\phi}$.
\end{center} 
%%%%%%%%%%%%%%(end of Figs.3,4)%%%%%%%%%%%  

 The IR divergence is regularized by introducing an  
infinitesimal mass, $\lam$, for the traceless mode as 
$1/k^4 \arr 1/(k^2+\lam^2)^2$.  
After taking contractions of the tensor indices, we obtain the following 
contribution from the diagram (a) in Fig.~3:
\bba
   &&  \int \frac{d^D k}{(2\pi)^D} \phi_r(k)\phi_r(-k) 
         \biggl[ -\frac{1}{6}b_r t_r^2 
       \int \frac{d^D l}{(2\pi)^D} \frac{1}{l^4 ~((l+k)^2 +\lam^2 )^2} 
                  \nonumber \\ 
   && \quad \times
     \biggl\{ 6(l^2 k^6 +l^6 k^2) +24l^4 k^4 -16(\lk)(l^2 k^4 +l^4 k^2) 
                -20 (\lk)^2 l^2 k^2 
                    \nonumber \\ 
   && \qquad       
           -2 (\lk)^2 (l^4 +k^4) +8(\lk)^3(l^2+k^2) +8(\lk)^4   
                     \nonumber \\ 
   && \qquad   
         + \frac{4-D}{3D} \biggl( 
          -36 l^4 k^4 +24 (\lk) (l^2 k^4 +l^4 k^2) +40(\lk)^2 l^2 k^2 
                   \nonumber \\ 
   && \qquad\qquad  
                 -4(\lk)^2 (l^4 +k^4) -16(\lk)^3 (l^2 +k^2) -16(\lk)^4 
                 \biggr) \biggr\} \biggr] ~.                                
\eea
Integrating over $l$ in the brackets, 
we obtain the divergence in the form 
\bb
     \frac{2 b_r}{(4\pi)^2} ~k^4 \biggl[ 
        -3\tt_r^2 \biggl( \frac{1}{\beps} 
           -\log \frac{\lam^2}{\mu^2} +\frac{1}{6} 
            \biggr) \biggr] ~, \label{fig3a}
\ee
where $\frac{1}{\beps} =\frac{1}{\eps} -\gm +\log 4\pi$. 
Here, we also compute finite terms for two-loop computations given below. 
The non-local term, $\log k^2/\mu^2$, does not appear 
in the finite term.   

   Diagram (b) in Fig.~3 is rather simple. The two terms in the 
vertex ${\cal L}^{(4)}_{\rm int}$, shown in (\ref{vx4}), produce 
logarithmic divergences. The other terms, including derivatives of $h$, 
give vanishing contributions. Hence, we obtain the following divergence:
\bb
     \frac{2 b_r}{(4\pi)^2} ~k^4 \biggl[ 
        3\tt_r^2 \biggl( \frac{1}{\beps} 
           -\log \frac{\lam^2}{\mu^2} -\frac{5}{12} 
            \biggr) \biggr] ~. \label{fig3b}
\ee
Therefore, the UV divergences as well as the IR divergences cancel out  
in the sum of (\ref{fig3a}) and (\ref{fig3b}), and we obtain 
\bb
          Z_b Z_{\phi}=1  \label{zbzp}
\ee
at order $t_r^2$. 

  This result is related to the counterterm, $\Delta {\cal I}_{R^2}$. 
Since $\sq{\hg}{\hat R}^2$ produces the kinetic term $\s \Box^2 \s $ 
when the background field of the conformal mode 
$\hg_{\mu\nu}= \e^{2\s}\dl_{\mu\nu}$ is considered,  
background-metric independence for the conformal mode 
implies the relations  
$Z_a Z_{\s} =Z_b Z_{\s} =Z_c Z_{\s}$  
of the renormalization factors, where $Z_{\s}$ is the 
renormalization factor of the background conformal mode. 
Since the background metric is not renormalized such that $Z_{\s}=1$, 
we obtain   
\bb
     Z_{\phi} = 1 ~, \qquad Z_a =Z_b =Z_c ~.   
\ee  
Thus, result (\ref{zbzp}) leads to  
\bb
         Z_a = Z_b = Z_c = 1  \label{zabc}
\ee
at order $t_r^2$. 
Since we do not introduce a coupling for the conformal mode, 
this result implies that the terms of order $t_r^2$ in $P_{R^2}$, 
(\ref{pt3}), vanish, so that   
\bb
             l_1 = l_1^{\pp} =0 ~.  \label{lv}
\ee
Here, $l_1 =0$ is rather trivial, 
because only the one-loop diagrams contribute to order $t_r^2$, which is a  
consequence of the cubic and quartic self-interactions of $\phi$ being  
dropped, by condition (\ref{d}). 
The result (\ref{lv}) is evidence that an action $S$ satisfying 
the integrability condition~\cite{r} exists at $t_r \neq 0$.  
This is discussed in further detail in the next subsection. 

  The relation (\ref{zth}) to order $t_r^2$ implies that   
the sum of the vertex diagrams (a) through (e), shown in Fig.~5, 
makes a finite contribution. 
This is likely to be true, as in the case of the self-energy diagrams. 
However, because direct calculations would be quite tedious, 
we need software to take contractions of the tensor indices 
into consideration.   
%%%%%%%%%%%%%%%(diagrams Figs.5)%%%%%%%%%%%%
%Vertex diagarms a,b,c
\begin{center}
\begin{picture}(400,100)(0,-10)
\Line(60,10)(140,10)
\PhotonArc(100,10)(25,0,180){3}{8.5} 
\Vertex(75,10){1}
\Vertex(125,10){1}
\Photon(100,38)(100,61){3}{2.5}
\Vertex(100,38){1}
\Text(100,-3)[]{(a)} 

\Line(170,10)(230,10)
\PhotonArc(200,31)(18,22,382){3}{12}
\Vertex(200,10){1}
\Photon(200,52)(200,75){3}{2.5}
\Vertex(200,52){1}
\Text(200,-3)[]{(b)} 

\Line(260,37)(340,37)
\PhotonArc(300,37)(25,180,360){3}{8.5} 
\Vertex(275,37){1}
\Vertex(325,37){1}
\Photon(300,37)(300,60){3}{2.5}
\Vertex(300,37){1}
\Text(300,-3)[]{(c)} 
\end{picture}
\end{center}
%Vertex diagarms d,e.
\begin{center}
\begin{picture}(300,100)(0,-10)
\Line(60,37)(140,37)
\PhotonArc(100,37)(25,180,360){3}{8.5} 
\Vertex(75,37){1}
\Vertex(125,37){1}
\Photon(75,37)(75,60){-3}{2.5}
\Text(100,-3)[]{(d)} 

\Line(170,51)(230,51)
\PhotonArc(200,30)(18,22,382){3}{12}
\Vertex(200,51){1}
\Photon(200,51)(200,74){3}{2.5}
\Text(200,-3)[]{(e)} 
\end{picture} \\ 
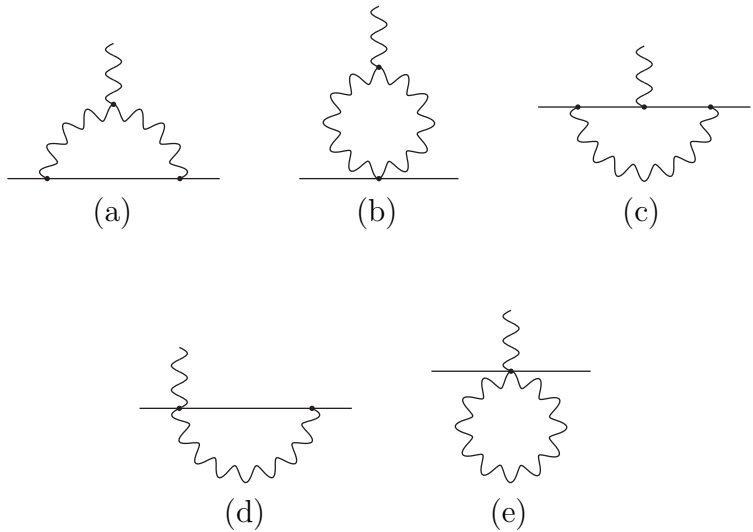
Fig. 5: One-loop vertex diagrams to order $t_r^2$. 
\end{center}
%%%%%%%%%%%%%%%(end of Fig.5)%%%%%%%%

  Quantum corrections proportional to $\phi^3$ and $\phi^4$, whose terms   
are absent in the action ${\cal I}$, arise at order $t_r^4$. 
Replacing the wavy external lines of diagrams (a) and (b) in Fig.~5 
with lines, we obtain corrections to $\phi^3$ of order $t_r^4$. 
Their sum is also likely to be finite, as in the self-energy diagrams 
of $\phi$.  In the same way, the divergences proportional to $\phi^4$ 
may cancel out. 

  Hence, we see that, from (\ref{zth}) and (\ref{zbzp}), the vertices in  
the Wess-Zumino action are likely to be 
unrenormalized to order $t_r^2$, which implies that non-local 
terms related to such vertices do not appear, 
and thus the Wess-Zumino action does not change form.  

\subsection{Diffeomorphism invariance at two-loop} 
\noindent
  
  In this subsection we discuss the condition of diffeomorphism 
invariance at the two-loop level. To clarify the origin of 
the interactions, we divide the combined Lagrangian into three parts,  
as follows:
\bba
   && {\cal L}^C =2b \phi {\bar \Delta}_4 \phi 
                + \frac{1}{t^2} {\bar F}  
                + {\cal L}_A (A,\bg) ~, 
                        \\ 
   && {\cal L}^{CS} = a {\bar F} \phi + b {\bar G} \phi ~, 
                         \\ 
   && {\cal L}^{NC} = -\frac{2}{3} b \bBox \bR \phi 
                      +\frac{1}{36}(2a+2b+3c) \bR^2 ~.  
\eea
Here, ${\cal L}^C$ is  conformally invariant. 
Expanding ${\cal L}^{CS}$ produces conformally invariant 
vertices linear in $\phi$. On the other hand, 
the last one, ${\cal L}^{NC}$, yields non-conformally 
invariant vertices. 

  The regularized 1PI effective action is divided into three parts as   
\bb
      \Gamma_{reg}(\phi_r,\bg_r)= {\cal I}_r(\phi_r,\bg_r,A_r) 
              +V_{reg}(\phi_r, \bg_r) + W_{reg}(\bg_r,A_r) ~. 
\ee 
Here, finite corrections $V_{reg}$ and $W_{reg}$ come from 
loop diagrams. The former represents corrections to the Wess-Zumino 
action, and the latter represents corrections independent of $\phi$,    
which are related to the counterterms for the traceless mode, 
$\Delta {\cal I}_{F,~G,~R^2}$, and those for matter fields. 
{}From the sum of the results (\ref{fig3a}) 
and (\ref{fig3b}), we obtain $V_{reg} = -\frac{7}{4}t^2_r 2 b_r 
\phi {\bar \Delta}_4 \phi + \cdots$. 
As discussed in subsection 3.2, there is no non-local 
correction in this part. 
  
   Diffeomorphism invariance now implies that the 1PI regularized 
effective action  satisfies 
the equation $\dl \Gamma_{reg}=0$, or 
\bba
    \dl_{\om} W_{reg}(\bg_r,A_r) 
          &=& \frac{1}{(4\pi)^2} \int d^4 x \sq{\hg_r} \om  
              \biggl[   a_r (1+v_a) \biggl( {\bar F}_r 
              + \frac{2}{3}\bBox_r \bR_r \biggr) 
                   \nonumber  \\ 
          && \qquad\qquad\qquad\qquad 
             + b_r (1+v_b){\bar G}_r 
             + c_r  \bBox_r \bR_r \biggr]  ~,   
                   \label{di}
\eea 
where $\om$ is an infinitesimal value (\ref{om4d}).   
The r.h.s. of (\ref{di}) comes from the variation of 
the actions ${\cal I}_r$ and $V_{reg}$, 
where $v_a$ and $v_b$  denote  
contributions from $\dl V_{reg}$. While $v_b$ has  been 
computed to order $t_r^2$, with    
\bb
   v_b=-\frac{7}{4}t^2_r ~,   
            \label{vbc}
\ee
$v_a$ has not.   
The l.h.s. of (\ref{di}) corresponds to the anomalous contributions 
from the measures $U(\om)$. 
Here, $\dl W_{reg}=\dl_{\om} W_{reg}$, because $W_{reg}$ 
is not include the conformal mode $\phi$, so that   
decomposing $\dl \bg_{\mu\nu}$ into 
$\dl_c \bg_{\mu\nu} = \bg_{\mu\lam}\bnabla_{\nu}\xi^{\lam} 
+\bg_{\nu\lam}\bnabla_{\mu}\xi^{\lam}$  
and $\dl_{\om}\bg_{\mu\nu}=2\om \bg_{\mu\nu}$,  
$W_{reg}$ is invariant under $\dl_c\bg_{\mu\nu}$. 
As seen below, (\ref{di}) uniquely determines 
coefficients $a_r$, $b_r$ and $c_r$. 

  The background-metric independence of the traceless mode implies 
that $\hg$ and $h$ always appear in the combination $\bg=\hg\e^h$.  
Therefore, it is enough to know the $\hg$ dependence of $W_{reg}$ to 
solve the condition of diffeomorphism invariance.   

  In the following arguments we consider only the $\hg$ dependence 
of $W_{reg}$. We here omit tensor indices 
for simplicity. The Weyl tensor is then described as 
${\hat C}$ and ${\hat C}^2={\hat F}$, 
where the subscript $r$ can be dropped, because 
$\hg_r =\hg$. The Euler density is also written as 
${\hat G}={\hat B}^2$. 

  Let us consider the $W_{reg}$ part, which is related to three  
counterterms, $\Delta {\cal I}_F$, $\Delta {\cal I}_G$ and 
$\Delta {\cal I}_{R^2}$.
First, we discuss the one-loop case~\cite{duff}. 
At this order, only the 
conformally invariant Lagrangian, ${\cal L}^C$, contributes to 
the effective action. {}From a dimensional analysis, 
the one-loop effective action should have the form~\cite{ddi}  
\bba
     W^{(1)} &=&   
           \frac{\tt_r^2}{2} \hh_r k^4 
           \biggl( \frac{k^2}{\mu^2} \biggr)^{-\eps} \biggl( 
            \frac{v}{\eps} + v^{\pp} \biggr) \hh_r + o(\eps)
               \nonumber       \\ 
           &\arr&  v_f \biggl( \frac{1}{\eps}{\hat F} 
                - {\hat C} \log (-\hBox/\mu^2){\hat C} \biggr) 
                      + v_f^{\pp} {\hat F}  
               \nonumber \\ 
           && + v_g \biggl( \frac{1}{\eps}{\hat G} 
                - {\hat B} \log (-\hBox/\mu^2){\hat B} \biggr) 
                      + v_g^{\pp} {\hat G} + o(\eps) ~,
                          \label{eff1}
\eea
where $v$ and $v^{\pp}$ represent    
$v_f$ and $v_g$ and $v_f^{\pp}$ and $v_g^{\pp}$, related to 
the conformally invariant counterterms $\Delta {\cal I}_F$ and 
$\Delta {\cal I}_G$, respectively. 
The IR cutoff mass is canceled out. Going to the second line, 
$(k^2)^{-\eps}= 1-\eps \log k^2 + \frac{1}{2!} \eps^2 \log^2 k^2 + \cdots$ 
is used. The coefficients $v_f$ and $v_g$ have been computed as 
$v_f=f_0$ and $v_g=g_0$ in (\ref{fg}). 
Hence, the counterterm can be written as  
\bb
         \Delta W^{(1)} = - \frac{f_0}{\eps} {\hat F} 
                          - \frac{g_0}{\eps} {\hat G} ~, 
                  \label{ct1}
\ee
and the regularized effective action is described as  
\bba
      W^{(1)}_{reg} &=& W^{(1)} + \Delta W^{(1)} 
             \nonumber     \\ 
         &=& -f_0 {\hat C} \log (-\hBox/\mu^2){\hat C}  
             -g_0 {\hat B} \log (-\hBox/\mu^2){\hat B} 
             + v_f^{\pp} {\hat F} + v_g^{\pp} {\hat G} ~.  
\eea

  The anomalous contribution is given by the variation of 
regularized effective action with respect to the conformal change as  
\bb
     \dl_{\om} W^{(1)}_{reg} = a_0 A + b_0 A^{\pp} ~,
          \label{reg1}
\ee
with
\bb
        a_0 =2f_0 ~, \qquad b_0=2g_0 ~, 
\ee
where $A= \om ( {\hat F} +(2/3)\hBox \hR)$ and $A^{\pp}= \om {\hat G}$. 
To derive this expression we use the following variational 
formulae~\cite{duff}: 
\bba
   &&  \dl_{\om}{\hat F} = - 2\eps A ~, \qquad  
     \dl_{\om}({\hat C} \log(-\hBox) {\hat C} )
     = - 2A +o(\eps) ~, 
                      \\ 
   &&  \dl_{\om}{\hat G} = - 2\eps A^{\pp} ~, \qquad  
     \dl_{\om}({\hat B} \log(-\hBox) {\hat B} )
     = - 2A^{\pp} +o(\eps) ~.
\eea
Here, note that using the formulae, we obtain the equation 
$\dl_{\om} W^{(1)}=0$,\footnote{%%%%%%%%%%%%%(footnote 7)%%%%%%
Precisely, we have $\dl_{\om} W^{(1)}= o(\eps)$, which reflects 
the fact that the conformal invariance of ${\cal L}^C$ is violated 
by $o(\eps)$ in $D$ dimensions.
}  %%%%%%%%(end of footnote 7)%%%%%%%%%%  
so that $\dl_{\om} W^{(1)}_{reg} = \dl_{\om} \Delta W^{(1)}$, 
as used in section 2. 

 Hence, substituting the value (\ref{reg1}) 
into the l.h.s. of (\ref{di}), 
we obtain the values of $a_r$ and $b_r$ 
at $t_r =0$ as $a_0$ and $b_0$. Here, note that 
$a_0$ is always negative and $b_0$ is positive, and their sum,     
$d_0=\frac{1}{18}(a_0+b_0)$, is positive.   

  The terms related to the non-conformally invariant counterterm, 
$\Delta {\cal I}_{R^2}$, in the regularized effective action in general 
have the forms  ${\hat R} \log (-\hBox/\mu^2) {\hat R}$ and ${\hat R}^2$. 
The term ${\hat R} \log (-\hBox/\mu^2) {\hat R}$ 
breaks the condition of diffeomorphism invariance, (\ref{di}), 
and therefore it must vanish. This condition is now equivalent to 
the vanishing of $l_0$, (\ref{l0v}).     
This implies that in order for the theory to be diffeomorphism invariant, 
those parts of the action contributing to the one-loop 
effective action $W^{(1)}$ should be conformally invariant. 
This is now ensured by condition (\ref{d}) for the gravitational fields.  
Matter fields must be conformally coupled to gravity;  
only in this case does 
the theory become diffeomorphism invariant. 
Since the variation of the $\hR^2$ term is proportional to 
$\hBox \hR$ (\ref{vosr}), this term determines the value of $c_r$. 
In the one-loop case, however, the conformal invariance is exact, 
and hence such a non-conformally invariant term vanishes. 
Thus, the value of $c_r$ at $t_r =0$  vanishes.  
  
  Let us analyze the effective action of order $t_r^2$ 
(or order $t_r^4$ for $Z_{\hh}$).  At this order, the non-conformally  
invariant terms in the action contribute to it.  
The effective action of order $t_r^2$ is divided into three parts as
\bb
      W^{(2)} = W^{(2-loop)} + W^{(counter)} + W^{(1-loop)}  ~, 
\ee
where $W^{(2-loop)}$ represents contributions of two-loop diagrams, 
which are derived from the conformally invariant Lagrangian ${\cal L}^C$. 
The contribution from associated diagrams with a one-loop 
counterterm insertion is represented by $W^{(counter)}$. 
Furthermore, there are contributions of order $t_r^2$, 
but they are one-loop diagrams described by $W^{(1-loop)}$, 
which include the vertices derived from ${\cal L}^{CS}$ 
and ${\cal L}^{NC}$.

  Since two-loop diagrams have two internal-momentum 
integrals, from a simple dimensional analysis, 
we find that $W^{(2-loop)}$ has the following form: 
\bba
     W^{(2-loop)} &=& \frac{\tt_r^4}{2} \hh_r k^4 \biggl[ 
             \biggl( \frac{k^2}{\mu^2} \biggr)^{-2\eps}  \biggl( 
               \frac{x}{\eps^2} + \frac{x^{\pp}}{\eps} 
                + x^{\pp\pp} \biggr) 
                  \nonumber \\     
         && \qquad\quad~
             + \biggl( \frac{k^2}{\mu^2} \biggr)^{-\eps}
             \biggl( \frac{\lam^2}{\mu^2} \biggr)^{-\eps}  \biggl( 
               \frac{y}{\eps^2} + \frac{y^{\pp}}{\eps} 
               + y^{\pp\pp} \biggr) 
                  \nonumber \\                                  
         && \qquad\quad~ 
            + \biggl( \frac{\lam^2}{\mu^2} \biggr)^{-2\eps}  \biggl( 
           \frac{z}{\eps^2} + \frac{z^{\pp}}{\eps} 
               + z^{\pp\pp} \biggr) \biggr] \hh_r ~. 
\eea
Here, the dependences of the momentum and the IR cutoff mass, 
$(k^2)^{-2\eps}$, $(k^2 \lam^2)^{-\eps}$ 
and $(\lam^2)^{-2\eps}$, in each term are the features of two-loop 
diagrams. On the other hand, since the associated counterterm insertion 
diagrams are one-loop, $W^{(counter)}$ should have the following 
form: 
\bb
    W^{(counter)} = \frac{\tt_r^4}{2} \hh_r 
             \biggl( -\frac{f_0}{\eps} \biggr) k^4 \biggl[ 
             \biggl( \frac{k^2}{\mu^2} \biggr)^{-\eps}   \biggl( 
                \frac{d}{\eps} + d^{\pp} \biggr) 
              +\biggl( \frac{\lam^2}{\mu^2} \biggr)^{-\eps}  \biggl( 
                \frac{e}{\eps} + e^{\pp} \biggr) 
                \biggr] \hh_r ~. 
\ee

  Since the contributions of two-loop diagrams come from the conformally 
invariant Lagrangian ${\cal L}^C$, the related two-loop counterterms 
becomes conformally invariant.  
The conditions of renormalizability and IR finiteness give constraints 
on the coefficients. From the condition that the non-local divergence, 
$\frac{1}{\eps}\log \frac{k^2}{\mu^2}$, cancels in the 
sum of $W^{(2-loop)}$ and $W^{(counter)}$, 
we obtain the constraint $-2x-y +f_0 d=0$. 
The cancellation of the IR divergences gives the constraints $y=z=e=0$ 
and $y^{\pp}+2z^{\pp}-f_0 e^{\pp}=0$.  
Hence, we obtain the following expression: 
\bba 
    && W^{(2-loop)} + W^{(counter)}
                   \nonumber \\ 
    && \arr \tt_r^2 \biggl[ 
      -  \frac{x_f}{\eps^2}{\hat F} -  \frac{x_g}{\eps^2}{\hat G} 
           +(x_f^{\pp}-z_f^{\pp}-f_0 d_f^{\pp})\frac{1}{\eps} {\hat F}
        +(x_g^{\pp}-z_g^{\pp}-f_0 d_g^{\pp})\frac{1}{\eps} {\hat G}
                           \nonumber \\ 
    && \qquad~
        + x_f {\hat C} \log^2 (-\hBox/\mu^2) {\hat C} 
        + x_g {\hat B} \log^2 (-\hBox/\mu^2) {\hat B} 
                \nonumber \\ 
    && \qquad~
        - (2x_f^{\pp}+y_f^{\pp}-f_0 d_f^{\pp}) 
         {\hat C} \log (-\hBox/\mu^2) {\hat C}  
         - (2x_g^{\pp}+y_g^{\pp}-f_0 d_g^{\pp}) 
         {\hat B} \log (-\hBox/\mu^2) {\hat B}  
                           \nonumber \\ 
    && \qquad~ 
        +(x_f^{\pp\pp}+y_f^{\pp\pp}+z_f^{\pp\pp})  {\hat F} 
        +(x_g^{\pp\pp}+y_g^{\pp\pp}+z_g^{\pp\pp})  {\hat G} 
               \biggr]~. 
               \label{ww2c}
\eea

  The diffeomorphism invariance (\ref{di}) requires  
$\dl_{\om} W_{reg}^{(2)}$ to be proportional to the anomaly forms $A$ 
and $A^{\pp}$. This implies that the finite term including the square 
of the logarithm should vanish, so that 
\bb
            x_f=x_g=0  \label{x} ~,  
\ee 
which is equivalent to the vanishing of the double pole 
in the two-loop counterterm. Thus, the renormalization factor of the 
background traceless mode, $Z_{\hh}$, does not have double poles of 
order $t_r^4$. On the other hand, using the relation between the 
renormalization factors, $Z_t Z_{\hh}^{1/2}=1$ in (\ref{zth}), and the 
finiteness of the $\beta$-function, one can easily prove that 
$Z_{\hh}$ does not have double poles of 
order $t_r^4$.\footnote{%%%%%%%%%%%%%(footnote 8)%%%%%%%%%
There is an analogy in QED, where the renormalization factor, defined by 
$e=Z_e e_r$, satisfies the relation $Z_e Z^{1/2}_3 =1$ as a result of 
the Ward-Takahashi identity $Z_1=Z_2$. The regularized 1PI vacuum 
polarization does not have $\log^n (-\Box/\mu^2)$ 
with $n \geq 2$. (See, for instance, section 8-4-4 in Ref.\cite{iz}.)  
} %%%%%%%%%%%%(end of footnote 8)%%%%% 
Thus, the relation $Z_t Z_{\hh}^{1/2}=1$ guarantees that we can impose the 
condition of diffeomorphism invariance at the two-loop level.  

  Furthermore, the relation $Z_t = Z_{\hh}^{-1/2}$ implies that 
the diagrams with counterterm insertions at vertices and  at 
propagators cancel each other, and hence 
$W^{(counter)}$ vanishes.\footnote{%%%%%%%%(footnote 9)%%%%%%%%%
Precisely, gauge-fixed renormalization insertion diagrams may be 
necessary~\cite{ab}.
} %%%%%%%%%%(end of footnote 9)%%%%%%% 
The constraint on the IR divergences now 
becomes  $y_{f,g}^{\pp}+2z_{f,g}^{\pp}=0$. 
Therefore, expression (\ref{ww2c}) reduces to 
the form~\footnote{ %%%%%%(footnote 10)%%
Note that, from a dimensional analysis, we cannot say that 
$\dl_{\om} W^{(2-loop)}$ vanishes. 
Rather, it satisfies $\mu \frac{\pd}{\pd\mu} W^{(2-loop)}=0$.
} %%%%%%%%%%%(end of footnote 10)%%%%%%% 
\bba 
    &&  \tt_r^2 \biggl[  
           (x_f^{\pp}-z_f^{\pp})\frac{1}{\eps} {\hat F}
        +(x_g^{\pp}-z_g^{\pp})\frac{1}{\eps} {\hat G}
                           \nonumber \\ 
    && \quad 
        - 2(x_f^{\pp}-z_f^{\pp}) 
         {\hat C} \log (-\hBox/\mu^2) {\hat C}  
         - 2(x_g^{\pp}-z_g^{\pp}) 
         {\hat B} \log (-\hBox/\mu^2) {\hat B}  
                           \nonumber \\ 
    && \quad 
        +(x_f^{\pp\pp}+y_f^{\pp\pp}+z_f^{\pp\pp})  {\hat F} 
        +(x_g^{\pp\pp}+y_g^{\pp\pp}+z_g^{\pp\pp})  {\hat G} 
               \biggr]~. 
             \label{w2c}
\eea

  Next we consider the one-loop part of the effective action $W^{(1-loop)}$. 
Recall that it includes the interactions derived from the Lagrangians 
${\cal L}^{CS}$ and ${\cal L}^{NC}$. 
Since ${\cal L}^{NC}$ is not conformally invariant, we must consider here 
the non-conformally invariant counterterm $\Delta {\cal I}_{R^2}$ also. 
{}From a dimensional analysis, we obtain the form 
\bba
    W^{(1-loop)} &=& \frac{\tt_r^4}{2} \hh_r k^4 
             \biggl( \frac{k^2}{\mu^2} \biggr)^{-\eps} 
             \biggl( \frac{w}{\eps} + w^{\pp}  \biggr) \hh_r ~. 
                 \nonumber \\ 
     &\arr& \tt_r^2 \biggl[  
           w_f \biggl( \frac{1}{\eps}{\hat F} 
                - {\hat C} \log (-\hBox/\mu^2){\hat C} \biggr) 
                      + w^{\pp}_f {\hat F} 
                 \nonumber \\ 
       &&\quad
          + w_g \biggl( \frac{1}{\eps}{\hat G} 
                - {\hat B} \log (-\hBox/\mu^2){\hat B} \biggr) 
                      + w^{\pp}_g {\hat G}
                 \nonumber \\
       &&\quad
          + w_l \biggl( \frac{1}{\eps}\hR^2 
                - \hR \log (-\hBox/\mu^2) \hR \biggr) 
                      + w^{\pp}_l \hR^2 \biggr] ~.
                  \label{w1}
\eea

  Combining contributions (\ref{w2c}) and (\ref{w1}), 
the two-loop counterterm reads 
\bb
     \Delta W^{(2)} = - \frac{\tt_r^2}{\eps} \biggl( 
                         f_1^{\pp}{\hat F} + g_1^{\pp}{\hat G} 
                         + l_1^{\pp} \hR^2 \biggl) ~, 
\ee
with 
\bb                   
     f_1^{\pp} = x_f^{\pp} -z_f^{\pp} +w_f~, \qquad 
     g_1^{\pp} = x_g^{\pp} -z_g^{\pp} +w_g~, \qquad 
     l_1^{\pp} = w_l ~. 
\ee
The regularized effective action, 
$W_{reg}^{(2)}=W^{(2)}+\Delta W^{(2)}$, can then be expressed in the form
\bba
  W_{reg}^{(2)} &=& t_r^2 \biggl[  
       - f_1^{\pp\pp} {\hat C} \log (-\hBox/\mu^2) {\hat C}  
       - g_1^{\pp\pp} {\hat B} \log (-\hBox/\mu^2) {\hat B} 
                       \nonumber \\ 
    && \quad 
        - l_1^{\pp} \hR \log (-\hBox/\mu^2) \hR  
        + u_f {\hat F} + u_g {\hat G} + u_l \hR^2 
                \biggr] ~,       
\eea
with 
\bb                   
     f_1^{\pp\pp} = 2(x_f^{\pp} -z_f^{\pp}) +w_f~, \qquad 
     g_1^{\pp\pp} = 2(x_g^{\pp} -z_g^{\pp}) +w_g ~,  
\ee
where $u_f= x_f^{\pp\pp} +y_f^{\pp\pp}+z_f^{\pp\pp}+w_f^{\pp}$,  
$u_g= x_g^{\pp\pp} +y_g^{\pp\pp}+z_g^{\pp\pp}+w_g^{\pp}$ and 
$u_l=w_l^{\pp}$. 
  
  The requirement of diffeomorphism invariance is now equivalent 
to the requirement that the non-local term 
${\hat R} \log (-\hBox/\mu^2) {\hat R}$  
vanishes, or $l^{\pp}_1=0$. 
This has been indirectly verified by computing the self-energy 
diagrams of the conformal mode to order $t_r^2$ in subsection 3.2. 
{}From the viewpoint of the vacuum polarization of $\hh_{\mu\nu}$, 
this implies that the divergences contributing to $l^{\pp}_1$ in the 
one-loop diagrams, which include the vertices coming from the 
non-conformally invariant Lagrangian, ${\cal L}^{NC}$, cancel out. 
As discussed at the end of this subsection, this is consistent with 
the result in Ref.\cite{hs}. 

  Furthermore, the non-local term  
$\phi \hBox^2 \log (-\hBox/\mu^2) \phi$,  
which changes the form of the Wess-Zumino action, does not appear 
in the regularized effective action at order $t_r^2$. 
The background-metric independence of the 
conformal mode implies that the vanishing of this term  
is equivalent to the vanishing of the non-local term  
$\hR \log (-\hBox/\mu^2) \hR$. 

   The anomalous contribution at the two-loop level is now given by
\bb
       \dl_{\om} W_{reg}^{(2)} 
       = t_r^2 \Bigl( a_1 A + b_1 A^{\pp} 
                            + c_1 \om \hBox \hR \Bigr) ~,  
\ee 
with 
\bb
    a_1= 2 f_1^{\pp\pp} ~, \qquad b_1= 2 g_1^{\pp\pp} ~, \qquad 
    c_1= -12u_l ~. 
\ee    
Combining the one-loop results, (\ref{di}) is now 
solved as 
\bba
      a_r(1+v_a) &=& a_0 + a_1 t_r^2 + \cdots ~, 
              \nonumber  \\ 
      b_r (1+v_b) &=& b_0 +  b_1  t_r^2 
              +\cdots  ~,      
             \nonumber \\ 
      c_r  &=& c_1 t_r^2 + \cdots ~.
             \label{abc}
\eea  
The coefficients $a_1$, $b_1$ and $c_1$ are 
generally functions of the couplings $a_r$ and $b_r$, 
while $a_0$ and $b_0$ are given by the constants. 
Hence, the BMI condition can be solved perturbatively. 
Substituting the value of $v_b$ from (\ref{vbc}), we then obtain 
\bb 
  b_r=b_0 +\biggl( b_1+\frac{7}{4}b_0 \biggr) t^2_r + \cdots ~,  
\ee  
where $b_1$ is now a function of $a_0$ and $b_0$. 
This equation is used in subsection 3.4.
 
  At higher order, the conditions of diffeomorphism invariance are 
expressed as the vanishing of the non-local terms: 
${\hat C} \log^n (-\hBox/\mu^2) {\hat C}$ $(n \geq 2)$, 
${\hat B} \log^n (-\hBox/\mu^2) {\hat B}$ $(n \geq 2)$ and 
$\hR \log^n (-\hBox/\mu^2) \hR$ $(n \geq 1)$. 
Each term with $n \geq 2$ is likely to vanish to all orders, as in gauge 
theories. This is ensured by relation (\ref{zth}). 
The non-trivial condition is the vanishing of the term 
$\hR \log (-\hBox/\mu^2) \hR$, 
which is the generalized form of the integrability condition discussed 
by Riegert, Fradkin and Tseytlin~\cite{r}. 
This condition does not imply that $\Delta {\cal I}_{R^2}$ 
vanishes at higher order.  
This condition may be realized when we substitute the values $a_r$, $b_r$ 
and $c_r$ from (\ref{abc}) into the expression of the regularized 
effective action. 

 This seems to be a trick of dimensional regularization. If there is a  
regularization defined exactly in 4 dimensions, like the Pauli-Villars 
method, one would be able to compute the effective action  substituting 
the values of the constants $a_n$, $b_n$ and $c_n$ required order by 
order into the bare action at the begining. 
Then, such a trick will not be  necessary, and  
we could expect that all divergences will be 
renormalized into the coupling $t$ and  matter fields 
and other dimensional constants, and that the Wess-Zumino action 
would not change form. 

  The difficulty to prove the renormalizability of our model 
originates from the fact that to preserve diffeomorphism invariance, 
the Wess-Zumino action should not change form, while the symmetry 
to preserve the form of the Wess-Zumino action would be nothing but  
diffeomorphism invariance itself. What can be done seems to be merely 
to check for any self-consistencies of the theory. 
In this paper we have shown that our model 
can preserve diffeomorphism invariance self-consistently 
up to order $t_r^2$ (or order $t_r^4$ for $Z_{\hh}$). 
Since an extension to higher order would be  
unique due to the background-metric independence for the traceless mode,  
it is expected that our model preserves  
diffeomorphism invariance to all orders. 
   
   Here we comment on studies by Kawai, Kitazawa and Ninomiya~\cite{kkn}. 
While we apply dimensional regularization to the theory 
${\cal I}=S+I$ (\ref{zz}), defined on the metric $\bg$, 
it seems that, as far as using dimensional regularization, one could 
apply it to the theory (\ref{zzz}), defined on the metric $g$, 
taking great care with the conformal mode dependence, as discussed by them. 
In their method it is expected that the coefficients $a_r$, $b_r$ 
and $c_r$ are automatically determined, 
while our method has the advantage that 
conformal mode dependence is manifest, and hence 
there is an analogy to 2D quantum gravity.  
The relation between the two methods is unclear at present. 
However, it seems that the two methods might be the same 
in the limit $\eps \arr 0$, 
at least up to order $t_r^2$, because at this order the divergences of 
the self-energy diagrams of the conformal mode cancel out for arbitrary 
values of $a_r$, $b_r$ and $c_r$.  Detailed analyses of this matter 
seem to be important to prove renormalizability to all orders.
  
  To determine the values $a_1$, $b_1$ and $c_1$, it seems that we must  
carry out  full two-loop calculations. 
On the other hand, there is a naive argument that the 
regulator of the measure is given by the bilinear term of quantum fields 
in the combined Lagranigan,  
${\cal L} = \half \Phi^t {\cal K} \Phi + o(\Phi^3)$, 
where $\Phi^t =(\phi, h^{\mu}_{~\nu}, A_{\mu})$. 
Then, using the special property of fourth order operators, we obtain 
an annomalous contribution formally expressed in the form~\cite{hs}   
\bb 
    - 2 Tr \Bigl( \dl \om \e^{-\varepsilon {\cal K}} \Bigr) ~,  
              \label{trk}
\ee 
where $\varepsilon$ is a cut-off. This is essentially a one-loop 
contribution, and hence it determines the values of $w_f$ and $w_g$. 
The proof in Ref.\cite{hs} of the Wess-Zumino condition 
for fourth order operators now implies that $w_l ~(=l_1^{\pp}) = 0$, 
which is consistent with what we mentioned above.  
Here, it is worth commenting on our previous work. 
In the proof given in section 3 in Ref.\cite{hs}, 
it was assumed that an action $S$ satisfying the 
integrability condition~\cite{r} exists. 
Therefore, the proof implies that, 
if it does exist, the variation of the part of the Wess-Zumino action $S$ 
related to one-loop diagrams is given by (\ref{trk}).  
Applying the proof to our model, existence at 
$t = 0$ is guaranteed by condition (\ref{d}), but not for $t \neq 0$. 
However, since the $\hg$ dependence of $S$ is determined by 
the background-metric independence of the conformal mode, 
the $h$ dependence of $S$ is uniquely determined by the background-metric 
independence of the traceless mode, as mentioned in the Introduction. 
Thus, the existence of $S$ is guaranteed 
in our model, so that $w_l=0$.        

  Here we conjecture that only the one-loop diagrams contribute to 
$\dl_{\om} W_{reg}$, and thus $a_1=2w_f$ and $b_1=2w_g$, 
because (\ref{trk}) is the most simple expression giving 
the Wess-Zumino action at all orders.
Using the generalized Schwinger-DeWitt technique~\cite{bv}, 
as in Ref.\cite{hs}, 
we obtain~\footnote{%%%%%%%%%%%%%%%(footnote 11)%%%%%%%%%%
In Ref.\cite{hs} we used the gauge-fixing 
condition that the conformal mode and the traceless mode are mixed. 
However, this is incorrect, because we did not introduce the 
coupling $t$ for the conformal mode. 
To ensure gauge independence, we here use the Feynman-type gauge 
$\zeta=1$ in (\ref{gfix}), and the  
coefficients $a$ and $b$ are re-calculated. 
Now, the off-diagonal term in the operator ${\cal K}$ becomes 
fourth order, but it can be managed in the perturbation by reducing 
the expression of $Tr \log {\cal K}$ to the formula given in Ref.\cite{bv} 
by using an identity like, for example   
$\hBox \hnabla_{\mu}V_{\nu}= \hnabla_{\mu}\hBox V_{\nu} 
+ 2\hR^{\a}_{~\nu\mu\b}\hnabla^{\b}V_{\a} 
+ \frac{1}{4} \hR \hnabla_{\mu}V_{\nu}$, 
repeatedly, where the constraints 
$\hR_{\mu\nu}=\frac{1}{4}\hg_{\mu\nu} \hR$ and $\hnabla_{\mu}\hR=0$ 
are imposed as in Ref.\cite{hs}, so that the coefficient $c_r$ cannot 
be determined.        
}%%%%%%%%%%%%%%(end of footnote 11)%%%%%%%%% 
\bba
    2w_f &=& -  \fr{18a^2_0 +124 a_0 b_0 +109 b^2_0}{72 b_0} ~,
                    \\ 
    2w_g &=&   \fr{14}{9} a_0 + \fr{115}{72} b_0  ~.  
               \label{b} 
\eea
Here, $u_l$ is not calculated.

\subsection{The $t_r^2$-order corrections of the cosmological constant} 
\noindent
 
  Let us compute the quantum corrections of the cosmological constant, 
$\Lam \e^{\a\phi}$. In 4 dimensions, the constant $\a$ should satisfy 
the equation~\cite{am} 
\bb
        \a = {\rm dim}[\Lam]= 4+\gm_{\Lam} ~, 
               \label{cos4}
\ee 
where $\gm_{\Lam}$ is the anomalous dimension. 
Here, we calculate $\gm_{\Lam}$ up to order $t_r^2$. 
The contribution to zeroth order of $t_r$ is given by the one-loop 
diagram in Fig.~1, as in the two-dimensional model. 
The diagrams of order $t_r^2$ contributing to the renormalization 
factor proportional to $\a^2$ are the two 2-loop diagrams, (a) and (b),   
and the 1-loop diagram, (c), in Fig.~6. Unlike the two-dimensional model, 
there are contributions proportional 
to $\a^3$ and $\a^4$. They are given by diagrams (a) and 
(b) in Fig.~7, respectively. 
%%%%%%%%%%%%%%%(diagrams Figs.6,7)%%%%%%%%%%
%Two-loop corrections of cos. const. (4 dim.)
\begin{center}
\begin{picture}(300,120)(0,-10)
\PhotonArc(50,70)(18,-28,208){3}{10.5}
\Vertex(33,61){1}
\Vertex(67,61){1}
\CArc(50,50)(20,0,360)
\Text(52,12)[]{$\cdots$}
\Line(38,10)(50,30)
\Line(42,9)(50,30)
\Line(62,10)(50,30)
\Vertex(50,30){1}
\Text(50,-3)[]{(a)} 

\PhotonArc(150,87)(18,0,360){3}{11}
\Vertex(150,66){1}
\CArc(150,48)(18,0,360)
\Text(152,12)[]{$\cdots$}
\Line(138,10)(150,30)
\Line(142,9)(150,30)
\Line(162,10)(150,30)
\Vertex(150,30){1}
\Text(150,-3)[]{(b)} 

\PhotonArc(250,52)(22,0,180){3}{6.5}
\Vertex(228,52){1}
\Vertex(272,52){1}
\CArc(250,52)(22,180,360)
\Text(252,12)[]{$\cdots$}
\Line(238,10)(250,30)
\Line(242,9)(250,30)
\Line(262,10)(250,30)
\Vertex(250,30){1}
\Text(250,-3)[]{(c)} 
\end{picture} \\ 
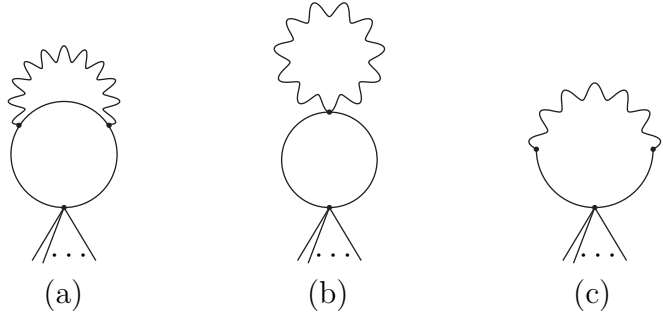
Fig. 6: Contributions of order $t_r^2$ proportional to $\a^2$. 
\end{center}
%(alpha)^3,4 corrections of cos. const.(4 dim.)
\begin{center}
\begin{picture}(300,120)(0,-10)
\PhotonArc(100,52)(22,0,180){3}{6.5}
\Vertex(78,52){1}
\Vertex(122,52){1}
\CArc(100,52)(22,180,270)
\Oval(111,41)(15,10)(-45)
\Text(102,12)[]{$\cdots$}
\Line(88,10)(100,30)
\Line(92,9)(100,30)
\Line(112,10)(100,30)
\Vertex(100,30){1}
\Text(100,-3)[]{(a)} 

\Oval(187,44)(15,10)(45) \Vertex(177,56){1} 
\Oval(213,44)(15,10)(-45) \Vertex(223,56){1} 
\PhotonArc(200,55)(22,0,180){3}{6.5}
\Vertex(200,38){1} 
\Text(202,12)[]{$\cdots$}
\Line(188,10)(200,38)
\Line(192,9)(200,38)
\Line(212,10)(200,38)
\Text(200,-3)[]{(b)} 
\end{picture} \\ 
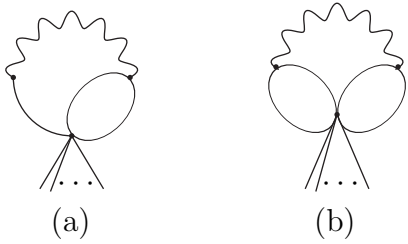
Fig. 7: Contributions of order $t_r^2$ proportional to $\a^3$ and $\a^4$. 
\end{center}
%%%%%%%%%%%%%%%%(end of Figs.6,7)%%%%%%%%%

  All of the contributions from Fig.~1, the sum of (a) and (b) in Fig.~6, 
and the other three ones, (c) in Fig.~6 and (a) and (b) in Fig.~7, are 
proportional to the tadpole-type integral $\int d^D k/k^4$. 
They thus yield only simple poles, which are summarized as 
\bb
      Z_{\Lam} -1 = - \biggl( \frac{1}{8\tb_r} 
                    +\frac{7}{32}\frac{\tt_r^2}{\tb_r} 
                    +\frac{\tt_r^2}{96} \biggr) \frac{\a^2}{\eps} 
                    + \frac{1}{1152}\frac{\tt_r^2}{\tb_r} 
                          \frac{\a^3}{\eps} 
                    -\frac{1}{3072} \frac{\tt_r^2}{\tb_r^2} 
                          \frac{\a^4}{\eps}   ~. 
\ee
The first term in the parentheses of the r.h.s. comes from the one-loop 
diagram in Fig.~1, and the second term comes from the sum of 
the two-loop diagrams, (a) and (b) in Fig.~6. 
The third  in the parentheses is the contribution from the one-loop 
(but order $t_r^2$) diagram (c) in Fig.~6. 
The last two terms on the r.h.s. come from (a) and (b) in Fig.~7, 
respectively. 

 There is no counterterm insertion diagrams, because all subdiagrams  
(namely, the sum of (a) and (b) in Fig.~3, 
the sum of (a) and (b) in Fig.~8, and (c) itself in Fig.~8) give 
finite contributions. 
Furthermore, diagrams (a) and (b) in Fig.~9 each gives a finite 
contribution. 
The finiteness of the sum of diagrams (a) and (b) in Fig.~8  
and of (c) in Fig.~8 is also evidence of renormalizability. 
%%%%%%%%%%%%%%%%%(diagrams Figs.8,9)%%%%%%%%%%%%
%Finite corrections of cos. const.I(4 dim.)
\begin{center}
\begin{picture}(300,100)(0,-10)
\Photon(18,76)(33,61){3}{3}
\Photon(82,76)(67,61){3}{3}
\Vertex(33,61){1}
\Vertex(67,61){1}
\CArc(50,50)(20,0,360)
\Text(52,12)[]{$\cdots$}
\Line(38,10)(50,30)
\Line(42,9)(50,30)
\Line(62,10)(50,30)
\Vertex(50,30){1}
\Text(50,-3)[]{(a)} 

\PhotonArc(150,87)(18,180,360){-3}{5.5}
\Vertex(150,66){1}
\CArc(150,48)(18,0,360)
\Text(152,12)[]{$\cdots$}
\Line(138,10)(150,30)
\Line(142,9)(150,30)
\Line(162,10)(150,30)
\Vertex(150,30){1}
\Text(150,-3)[]{(b)} 

\Photon(250,66)(250,87){-3}{2.5}
\Vertex(250,66){1}
\CArc(250,48)(18,0,360)
\Text(252,12)[]{$\cdots$}
\Line(238,10)(250,30)
\Line(242,9)(250,30)
\Line(262,10)(250,30)
\Vertex(250,30){1}
\Text(250,-3)[]{(c)} 
\end{picture}\\ 
Fig. 8: Diagrams which give finite contributions I. 
\end{center}
%Finite corrections of cos. const.II(4 dim.)
\begin{center}
\begin{picture}(300,100)(0,-10)
\PhotonArc(100,52)(22,0,180){3}{6.5}
\Vertex(78,52){1}
\Vertex(122,52){1}
\Line(63,52)(78,52)
\Line(122,52)(137,52)
\CArc(100,52)(22,180,360)
\Text(102,12)[]{$\cdots$}
\Line(88,10)(100,30)
\Line(92,9)(100,30)
\Line(112,10)(100,30)
\Vertex(100,30){1}
\Text(100,-3)[]{(a)} 

\PhotonArc(200,52)(22,0,180){3}{6.5}
\Vertex(178,52){1}
\Vertex(222,52){1}
\Line(163,52)(178,52)
\CArc(200,52)(22,180,270)
\Oval(211,41)(15,10)(-45)
\Text(202,12)[]{$\cdots$}
\Line(188,10)(200,30)
\Line(192,9)(200,30)
\Line(212,10)(200,30)
\Vertex(200,30){1}
\Text(200,-3)[]{(b)} 
\end{picture} \\ 
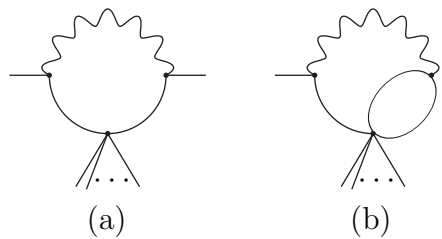
Fig. 9: Diagrams which give finite contributions II.
\end{center}
%%%%%%%%%%%%%%(end of Figs.8,9)%%%%%%%%%%

  The sum of the diagram of type (a) and associated 
counterterm-insertion diagrams (b) and (c) in Fig.~10  
does not yield  simple poles, where the diagram with the gray 
circle shown in Fig.~11 stands for the diagrams in Figs.~1, 6, 7, 8 and 9.  
If the gray part is a finite diagram, there is no counterterm   
of type (c) in Fig.~10, and the sum of the diagrams becomes finite.    
If the gray part is a divergent diagram, the sum of the diagrams yields 
only double poles, which only play the role to make an anomalous 
dimension finite.
%%%%%%%%%%%%%%%(diagrams Figs.10,11)%%%%%%%%%
%Double-pole or finite corrections of cos. const.(4 dim.)
\begin{center}
\begin{picture}(300,100)(0,-10)
\GOval(31,49)(22,16)(45){0.5}  
\Oval(63,44)(15,10)(-45)  
\Vertex(50,38){1} 
\Text(52,12)[]{$\cdots$}
\Line(38,10)(50,38)
\Line(42,9)(50,38)
\Line(62,10)(50,38)
\Text(50,-3)[]{(a)} 

\GCirc(150,50){20}{0.5}
\GCirc(150,30){4}{1}
\Line(147,33)(153,27)
\Line(147,27)(153,33)
\Text(152,12)[]{$\cdots$}
\Line(138,10)(150,26)
\Line(142,9)(150,26)
\Line(162,10)(150,26)
\Text(150,-3)[]{(b)} 

\CArc(250,45)(15,0,360)
\GCirc(250,30){4}{0.5}
\Line(247,33)(253,27)
\Line(247,27)(253,33)
\Text(252,12)[]{$\cdots$}
\Line(238,10)(250,26)
\Line(242,9)(250,26)
\Line(262,10)(250,26)
\Text(250,-3)[]{(c)} 
\end{picture} \\ 
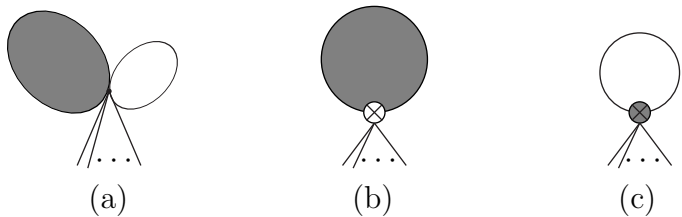
Fig. 10: Diagrams that give double poles or finite contributions.
\end{center}
%Corrections of cos. const. sumerized (4 dim.)
\begin{center}
\begin{picture}(100,90)(0,-10)
\GCirc(50,50){20}{0.5}
\Text(52,12)[]{$\cdots$}
\Line(38,10)(50,30)
\Line(42,9)(50,30)
\Line(62,10)(50,30)
\Vertex(50,30){1} 
\end{picture} \\ 
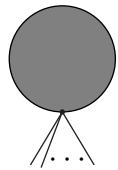
Fig. 11: Symbolical description of Fig. 1 and Figs. 6--9.
\end{center}
%%%%%%%%%%%%%%%(end of Figs.10,11)%%%%%%%%%% 

  Hence, we obtain the anomalous dimension up to order $t_r^2$ as 
\bb
         \gm_{\Lam} = \frac{1}{4}\frac{\a^2}{b_r} 
                       + t_r^2 \biggl( \frac{7}{8} \frac{\a^2}{b_r}  
                       + \frac{\a^2}{48}  
                      -\frac{1}{288} \frac{\a^3}{b_r} 
                      + \frac{1}{512} \frac{\a^4}{b_r^2} \biggr)  ~. 
\ee
The contributions proportional to $\a^n$ with 
$n>4$ appear at order $t_r^4$ and higher. 

   Until now, we have not specified the value of $b_r$, which is   
required by diffeomorphism invariance. Doing so, we obtain 
\bb
         \gm_{\Lam} = \frac{1}{4}\frac{\a^2}{b_0} 
                       + t_r^2 \biggl(  
                       -\frac{b_1}{4}\frac{\a^2}{b_0^2}  
                       +\frac{7}{16} \frac{\a^2}{b_0}  
                       + \frac{\a^2}{48}    
                      -\frac{1}{288} \frac{\a^3}{b_0} 
                      + \frac{1}{512} \frac{\a^4}{b_0^2} \biggr) ~. 
                        \label{gm}
\ee
Thus, (\ref{cos4}) becomes biquadratic at order $t_r^2$. 

   The cosmological constant should be physically real. 
Supposing that there is a real solution for the biquadratic equation, 
it can be solved perturbatively as $\a=\sum_n \a_n t_r^{2n}$. 
We then obtain 
\bba
   &&  \a_0 = 2b_0 \biggl( 1- \sq{1-\frac{4}{b_0}} \biggr)  ~,     
               \nonumber     \\ 
   &&  \a_1 = \frac{1}{\sq{1-\frac{4}{b_0}}}  
                     \biggl\{  \biggl(
                       -\frac{1}{4}\frac{b_1}{b_0^2}  
                       +\frac{7}{16} \frac{1}{b_0}  
                       + \frac{1}{48} \biggr) \a_0^2   
                      -\frac{1}{288} \frac{\a_0^3}{b_0} 
                      + \frac{1}{512} \frac{\a_0^4}{b_0^2} \biggr\}   ~,                                                                                                                                          \label{sol}
\eea
where we choose the solution of $\a_0$ so that $\a_0 \rightarrow 4$ 
in the classical limit, $b_0 \rightarrow \infty$. 

  Let us compare our model with the dynamical triangulation approach 
to 4D quantum gravity. 
In this approach there is no ambiguity choosing the 
measure, unlike with Regge's calculus. Furthermore, although  
this lattice model has parameters corresponding to 
the gravitational constant and the cosmological constant, 
there is no dimensionless coupling $t_r$, 
unlike in the usual lattice gauge theories. 
This indicates that, in continuum theory, the measure of a gravitational 
field including fourth order parts may be uniquely determined. 
Here, we estimate the strength of the coupling constant $t_r$ realized 
in  recent ``numerical experiments"~\cite{bbkptt}, 
which suggest that $\a$ is imaginary for $N_A=0$ but becomes real 
for $N_A \geq 1$. 

  Recall that $b_0$ is slightly greater than $4$ for a small 
value of $N_A$.  Thus, $\a_0$ is close to $8$. 
Therefore, the $\a^3$- and $\a^4$-terms and also the 
third term proportional to $\a^2$ in the parentheses of (\ref{gm}) are 
estimated as having a value on the order $o(1) \times t_r^2$, while 
the second term in the parentheses, which comes from the sum of the 
two-loop diagrams (a) and (b) in Fig.~6, is more than 10-times greater 
than them.   
Therefore, we neglect the terms proportional to $\a^3$ and $\a^4$.  
Then, the equation reduces to a quadratic form,  
and we obtain the following discriminant: 
\bb
     J = b_0 + \biggl( b_1- \frac{7}{4}b_0 \biggr) t_r^2 -4 ~. 
\ee 
The results of the simulations~\cite{bbkptt} can now be described as  
$J <0$ for $N_A =0$, but $J > 4$ for $N_A \geq 1$. 
If $b_1 =2 w_g$ is assumed,\footnote{%%%%%%%(footnote 12)%%%%%
Note that $2w_g$ is negative and that $|2w_g|$ is less than $b_0$ 
for small $N_A$. 
} %%%%%%%%%%(end of footnote 12)%%%%%%
there is a solution of the inequality, and we obtain the relation  
$0.025 < t_r^2 < 0.040$ for the coupling constant.  
If $b_1$ is neglected, the central value shifts by 
about $+0.01$.   
Expression (\ref{sol}) is valid in this region.

\section{Conclusions and Discussion} 
\setcounter{equation}{0}
\noindent

  In this paper we considered the problems of renormalizability 
and diffeomorphism invariance in a 4D quantum theory of gravity. 
We considered a perturbation theory for the traceless mode, where the 
dimensionless coupling constant $t$ was introduced,   
while the conformal mode was managed exactly. 
We found evidence that there is a model that satisfies the two conditions 
of renormalizability and diffeomorphism invariance simultaneously;  
it is uniquely determined to be model (\ref{cla}) with (\ref{d}) 
if we do not introduce fourth order fields, except for metric fields. 
In particular, the vanishing of $l_1$ and $l_1^{\pp}$ in (\ref{pt3}), 
which is required by the integrability condition discussed by 
Riegert, Fradkin and Tseytlin~\cite{r}, gives a breakthrough 
to the two-loop renormalizability of 4D quantum gravity. 

  Our conclusion seems to be plausible from the viewpoint of 
4D simplicial quantum geometry, because dynamical triangulation 
chooses a measure uniquely, 
and therefore the measure of the corresponding continuum theory 
may be determined uniquely. We expect that it may be given by 
the fourth order parts of our model.

  The fourth order model in 4 dimensions is quite similar to 
the second order model in two dimensions. 
We can rather easily compute the quantum 
corrections of the cosmological constant up to order $t_r^2$, 
as in the case of 2D quantum gravity, and we 
obtain the biquadratic equation (\ref{cos4}) with (\ref{gm}).  
Quantum corrections of the Einstein-Hilbert term are also quite 
important. Determining these is a future project. 

  In this paper we did not consider gauge interactions, 
such as QED and  Yang-Mills theories. 
If gauge couplings are introduced, one has to add a new coupling, 
$ \kappa \phi Tr(F_{\mu\nu}F^{\mu\nu})$, to the action $S$. 
The coefficient $\kappa$ then becomes a function of the gauge coupling,  
determined from the non-local logarithmic terms of the regularized 
vacuum polarizations of gauge fields.  

 To this point, we have focused only on the two fundamental conditions of 
renormalizability and diffeomorphism invariance. Theories satisfying 
these two conditions could be called ``quantum geometry" rather than  
``quantum gravity". To discuss quantum gravity, which should describe our 
universe, we need to define the $S$-matrix. 
%%%%%%%%%%%%%%%%(The S-matrix, Fig. 12)%%%%%%%%%% 
%The S-matrix
\begin{center}
\begin{picture}(300,120)(0,0)
\ArrowLine(25,80)(75,65)
\Photon(50,78)(75,90){3}{4}
\ArrowLine(25,20)(75,35) 
\Photon(50,22)(75,10){3}{4}
\GOval(100,50)(30,22)(0){0.5}
\Text(100,50)[]{fire ball}
\Photon(118,75)(168,95){3}{8}
\Photon(123,65)(173,85){3}{8}
\Photon(126,55)(176,75){3}{8}
\Text(230,85)[]{$\Biggl\}$ gravity emissions}
\ArrowLine(125,45)(175,35)\ArrowLine(120,30)(170,10) 
\Text(230,22)[]{$\Biggl\}$ matter emissions}
\end{picture} \\
%\begin{flushleft}
Fig. 12: 
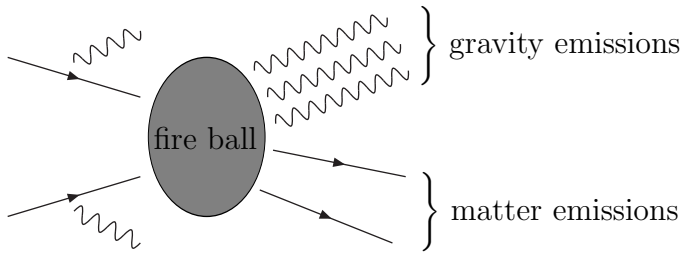
Image of the $S$-matrix of a fire ball/black hole formation $\quad$\\ 
and evaporation. The lines with arrows denote matter fields.  \\
The wavy lines are gravitational fields in the low-energy \\ 
asymptotic region.  In the fire-ball region, the fourth order \\ 
terms are turned on. 
$\qquad\qquad\qquad\qquad\qquad\qquad\qquad\quad$ 
%\end{flushleft}
\end{center}
%%%%%%%%%%%%%%%%(end of Fig. 12)%%%%%%%

  Recall, as computed in section 3.4, the cosmological constant 
receives non-trivial quantum corrections, because the propagators 
of the gravitational modes are proportional to $1/k^4$, and 
tadpole-type diagrams have logarithmic divergences.  
This is one difference between the fourth order theories 
and second order theories merely with negative-metric states.     
One can define diffeomorphism-invariant correlation functions 
in such a fully fluctuating geometry.  
However, when one attempts to define the $S$-matrix, one encounters 
the problem of how to define the asymptotic states 
for the gravitational modes. 
Since the states are always dressed by the conformal mode 
when the fourth order kinetic term of the conformal mode exists, 
it seems that one has to consider an unusual interaction picture 
in which the fourth order parts are turned off in the asymptotic region, 
as in a picture of a quantum black hole, where an asymptotically flat 
region exists (Fig.~12)~\cite{haw}. 
In this picture, negative-metric states in the fourth 
order terms required by diffeomorphism invariance do not appear 
in the low-energy asymptotic region, 
because there, gravity is ruled by the Einstein-Hilbert term. 
Hence, the negative-metric state is unstable and seems to be confined  
in the internal high-energy region, where the fourth order terms 
are turned on~\cite{lw,t,bhs}.   

  More directly, as a consequence of the background-metric independence, 
or diffeomorphism invariance, the presence of the $\phi {\bar F}$ term 
in the action implies that we can remove the Weyl term, ${\bar F}$, by 
shifting the conformal mode by a constant. 
Thus, although in this ``gauge" the theory becomes unrenormalizable, 
we can remove the negative-metric state for the traceless mode. 
This situation is somewhat reminiscent the argument of the $R_{\xi}$ gauge in 
spontaneously broken gauge theory~\cite{fls}. 

  Let us compute the degrees of freedom in this case. 
We must now treat the $\bR^2$ term and the Einstein-Hilbert term.  
Then, the gauge-fixing term $\chi^{\mu}\Box\chi_{\mu}$  
in (\ref{gfix}) is not necessary. 
Hence, $N_{\mu\nu}$ is proportional to  
$- \pd_{\mu}\pd_{\nu} +m^{\pp 2}\dl_{\mu\nu}$, 
where $m^{\pp 2}$ originates from the Einstein-Hilbert term, 
which is proportional to $m^2$.    
The ghost determinant is now given by 
$\det^{1/2} (- \pd_{\mu}\pd_{\nu}+m^{\pp 2}\dl_{\mu\nu}) 
\det M^{GH}_{\mu\nu}$, 
where $M^{GH}_{\mu\nu}$ is the usual second order ghost operator of 
diffeomorphism invariance defined by the relation 
$M^{GH}_{\mu\nu}\xi^{\nu}= \dl \chi_{\mu}|_{h=0}$. 
Note that $\det (- \pd_{\mu}\pd_{\nu}+m^{\pp 2}\dl_{\mu\nu}) 
= \det (-\Box+m^{\pp 2}) \vert_{\rm a ~scalar}$ 
(see, for instance, Ref.\cite{bv}), so that the number of ghost degrees 
of freedom is $4+4+1=9$. 
Hence, the total number of degrees of freedom becomes $2 \times 1+9-9=2$.  
Thus, the negative-metric state related to the conformal mode is 
also removed by ghosts. 
We expect that in the general case, the negative-metric states 
for the traceless mode may be confined, due to diffeomorphism invariance.

\vspace{5mm}

\begin{flushleft}
{\bf Acknowledgements}
\end{flushleft}

  I would like to thanks H. Kawai, Y. Kazama and 
T. Yoneya for helpful comments and suggestions on the manuscript. 
I also acknowledge H. Hagura, S. Horata, Y. Kitazawa, F. Sugino, 
A. Tsuchiya and T. Yoshikawa for valuable discussions.

\end{document}